\documentclass[%
superscriptaddress,
showpacs,
amsmath,amssymb,
aps,
pra,
longbibliography,
reprint,
twocolumn,
]{revtex4-1}

\usepackage{amsmath}
\usepackage{graphicx}
\usepackage{amssymb}
\usepackage{url}
\usepackage{booktabs}
\usepackage{mathrsfs}
\usepackage{color}
\usepackage[normalem]{ulem}
\definecolor{darkgreen}{rgb}{0.0, 0.2, 0.0}
\usepackage[colorlinks=true,linkcolor=blue,anchorcolor=blue,citecolor=blue,urlcolor=blue]{hyperref}

\begin{document}	

\title{Anomalous Autler-Townes Splitting in Resonant Multiphoton Ionization Driven by Bright Squeezed Vacuum}

\author{Xu Zhang}
\affiliation{School of Physics and Wuhan National Laboratory for Optoelectronics, Huazhong University of Science and Technology, Wuhan 430074, China}
\author{Liding Li}
\affiliation{School of Physics and Wuhan National Laboratory for Optoelectronics, Huazhong University of Science and Technology, Wuhan 430074, China}
\author{Yutong Deng}
\affiliation{School of Physics and Wuhan National Laboratory for Optoelectronics, Huazhong University of Science and Technology, Wuhan 430074, China}
\author{Xinyou Lv}
\affiliation{School of Physics and Wuhan National Laboratory for Optoelectronics, Huazhong University of Science and Technology, Wuhan 430074, China}

\author{Yang Li}
\email{liyang22@sjtu.edu.cn}
\affiliation{State Key Laboratory of Dark Matter Physics, Key Laboratory for Laser Plasmas (Ministry of Education) and School of Physics and Astronomy, Collaborative Innovation Center for IFSA (CICIFSA), Shanghai Jiao Tong University, Shanghai 200240, China}

\author{Marcelo F. Ciappina}
\email{marcelo.ciappina@gtiit.edu.cn}
\affiliation{Department of Physics, Technion--Israel Institute of Technology, Haifa 32000, Israel}
\affiliation{Department of Physics, Guangdong Technion--Israel Institute of Technology, Shantou 515063, Guangdong, China}
\affiliation{Guangdong Provincial Key Laboratory of Materials and Technologies for Energy Conversion, Guangdong Technion--Israel Institute of Technology, Shantou 515063, Guangdong, China}

\author{Peixiang Lu}
\email{lupeixiang@hust.edu.cn}
\affiliation{School of Physics and Wuhan National Laboratory for Optoelectronics, Huazhong University of Science and Technology, Wuhan 430074, China}

\author{Yueming Zhou}
\email{zhouymhust@hust.edu.cn}
\affiliation{School of Physics and Wuhan National Laboratory for Optoelectronics, Huazhong University of Science and Technology, Wuhan 430074, China}

\date{\today}
\begin{abstract}

Bright squeezed vacuum (BSV) light has a vanishing mean optical electric field yet can strongly enhance strong-field nonlinear responses beyond the conventional semiclassical paradigm.
Here we examine this scenario in the light-matter strong-coupling regime by investigating resonant multiphoton ionization of atoms driven by BSV, using a fully quantum treatment of both the electron and the field.
Our results show that the photoelectron energy spectrum exhibits an anomalous Autler-Townes splitting whose magnitude grows with the Above-threshold-ionization (ATI) order, rather than remaining essentially ATI-order independent as in the case of coherent driving. This behavior reflects a general scaling with the number of absorbed photons and originates from the broad photon-number fluctuations of the driving field together with the resulting electron-field entanglement. 
We further show that the BSV-induced enhancement of ionization yields evolves with intensity, crossing over from the $g^{(p+1)}$ limit to the $g^{(p)}$ limit as Rabi oscillations become established. These results identify a quantum regime of strong-field ionization governed by the interplay of photon statistics, nonlinear transitions, strong coupling, and nonseparable light-matter dynamics.

\end{abstract}	
\maketitle

The interaction of atoms and molecules with intense light is conventionally described within a semiclassical framework, in which the intense light acts as a prescribed classical wave and only the matter is treated quantum mechanically. 
This approach has successfully accounted for a wide range of highly nonlinear and non-perturbative phenomena, such as high harmonic generation (HHG) \cite{Paul2001,Agostini2004AttosecondLightPulses}, multiphoton ionization \cite{Bates1976,Charle1985}, strong-field above-threshold ionization (ATI) \cite{Agostini1979,Freeman1987}, and has shaped the standard understanding of strong-field dynamics in terms of laser intensity, polarization, frequency, and coherence.
Its central assumption, however, is that the driving field is regarded as an external control parameter rather than a dynamical quantum subsystem, so that its quantum properties are neglected.

In recent years, increasing attention has been devoted to the quantum nature of light in strong-field physics \cite{Tsatrafyllis2017High-order,Gombkoto2020HHGQuantizedField,Gorlach2020QuantumOpticalHHG,Foldi2021DescribingHHG,Lewenstein2021OpticalCatStates,RiveraDean2022LightMatterEntanglement,Stammer2022EntanglementMeasurementHHG,Stammer2022HighPhotonEntangled,Bhattacharya2023StrongLaserFieldPhysics,CruzRodriguez2024QuantumPhenomenaAttosecond,Hubenschmid2024OpticalTimeDomainTomography,Lange2024ElectronCorrelationNonclassicality,Pizzi2023LightEmissionManyBody,Theidel2024EvidenceQuantumOpticalHHG,Stammer2024EntanglementSqueezingHHG,RiveraDean2024SqueezedHHGExcitedAtoms,Gothelf2025HHGCrystalQuantumLight,Wang2025QuantumPathControlHHG,Wang2025HHGQuantumLightVonNeumann,ShiJunWang2025,Yi2025MassivelyEntangledBrightStates,Spasibko2017Multiphoton,Khalaf2023ComptonQuantumLight,EvenTzur2024SqueezedHighOrderHarmonics,Long2025HydrogenMolecularDissociation,Lemieux2025PhotonBunchingHHG,Gorlach2023HHGDrivenQuantumLight,Rasputnyi2024HighHarmonicBSV,Heimerl2024MultiphotonElectronEmission,Heimerl2025QuantumLightDrivesElectrons,EvenTzur2024MotionChargedParticles,Fang2023StrongFieldIonization,Liu2025AtomicDoubleIonizationQuantumLight}.
Particularly, progress in intense quantum-light generation has opened a route beyond the conventional semiclassical paradigm by making it possible to drive strong-field processes with highly nonclassical radiation \cite{Ishkhakov2012,Finger2015,CruzRodriguez2024QuantumPhenomenaAttosecond}.
Among such sources, bright squeezed vacuum (BSV) is especially appealing because it combines a large photon flux with giant photon-number fluctuations and strongly super-Poissonian statistics \cite{Schleich2001}. Recent experimental progress has further established BSV as a viable ultrafast quantum-light source: femtosecond BSV pulses with single-shot temporal characterization have now been demonstrated, showing that this highly fluctuating, zero-mean-field state can be produced and diagnosed on timescales relevant to strong-field and attosecond light-matter interactions \cite{Kern2026Optica}. These advances make BSV a timely platform for exploring nonlinear and strong-coupling regimes beyond the conventional coherent-state description of intense laser fields.
Previous studies have shown that quantum statistical properties of BSV can substantially reshape strong-field nonlinear responses \cite{Spasibko2017Multiphoton,Khalaf2023ComptonQuantumLight,EvenTzur2024SqueezedHighOrderHarmonics,Long2025HydrogenMolecularDissociation,Lemieux2025PhotonBunchingHHG,Gorlach2023HHGDrivenQuantumLight,Rasputnyi2024HighHarmonicBSV,Heimerl2024MultiphotonElectronEmission,Heimerl2025QuantumLightDrivesElectrons,EvenTzur2024MotionChargedParticles,Fang2023StrongFieldIonization,Liu2025AtomicDoubleIonizationQuantumLight}. For example, BSV-driven HHG exhibits extended cutoff energies \cite{Gorlach2023HHGDrivenQuantumLight}, enhanced harmonic yields and modified power-scaling laws \cite{Rasputnyi2024HighHarmonicBSV} compared to classical coherent driving; in quantum-light-driven strong-field ionization, the field statistics can also be imprinted onto the photoelectrons \cite{Heimerl2024MultiphotonElectronEmission,Heimerl2025QuantumLightDrivesElectrons}, thereby affecting electron trajectories \cite{EvenTzur2024MotionChargedParticles}, electron-wave-packet interference structures \cite{Fang2023StrongFieldIonization}, and ionization channels \cite{Liu2025AtomicDoubleIonizationQuantumLight}.
Most of these effects, however, have been explored in regimes where the light-matter interaction remains effectively weakly coupled. An essential question then naturally arises: how does quantum light reshape strong-field dynamics in the strong-coupling regime, where light and matter form hybrid dressed states and undergo reversible energy exchange? 

\begin{figure}[t]
		\centering
		\includegraphics[width=0.98\linewidth]{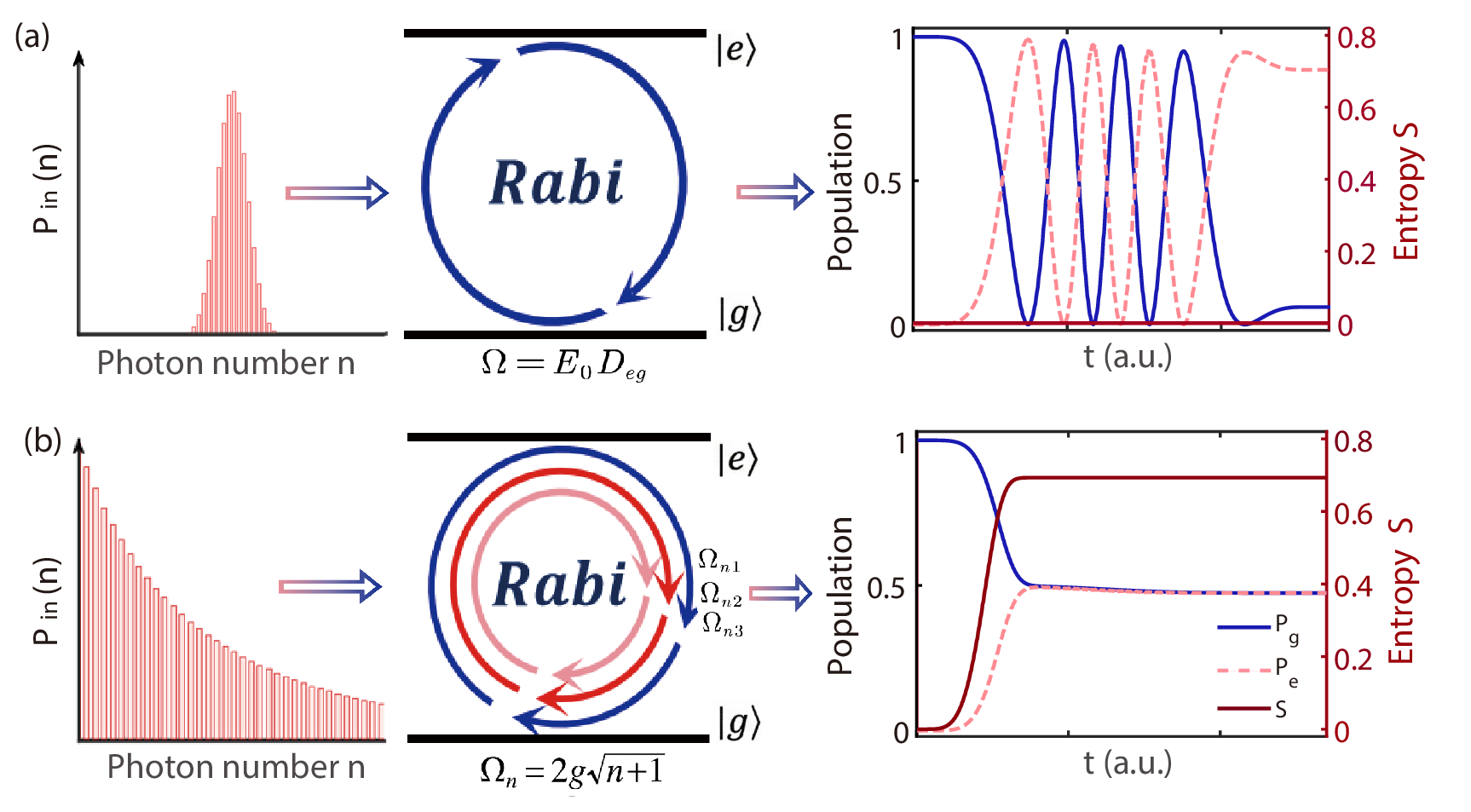}
        \caption{ Comparison of Rabi dynamics under intense coherent (classical) and BSV driving fields. (a) Intense coherent driving produces periodic Rabi oscillations at a single frequency (the populations of the corresponding atomic states are shown by blue and pink curves), with negligible atom-field entanglement (red curve). (b) BSV driving induces multi-frequency Rabi dynamics, leading to a rapid collapse of the oscillations (blue solid and pink dashed curves) and strong atom-field entanglement (quantified by the von Neumann entropy, red curve).}
		\label{Fig1}		
	\end{figure}

This question becomes particularly compelling in strong-field resonant regimes, where bound-state population transfer and Rabi oscillations build up \cite{Rabi1937,Knight1980,Walker1995,Sun2003,Wollenhaupt2003,Kaiser2013,Ciappina2015,Fushitani2015,Tumakov2019,LiWankai2021,Csehi2021,Jiang2021,Zhang2022,Nandi2022,Liao2022,olofsson2023,Wollenhaupt2023,Cui2023,Pan2023,Ishikawa2023,Nandi2024,Umarova2024,deLasHeras2025}. Under intense coherent driving, the dynamics follows the well-known semiclassical picture: the atomic population oscillates at a single Rabi frequency, the driving field can still be regarded as an effectively classical background that remains essentially unchanged during the interaction, and light-matter entanglement remains negligible throughout [Fig.~\ref{Fig1}(a)]. 
Under BSV driving, by contrast, the broad photon-number distribution causes different Fock components to evolve with different effective couplings. As a result, the bound-state populations undergo a rapid collapse, and the light-atom entanglement rises strongly [Fig.~\ref{Fig1}(b)]. In such a scenario, the semiclassical dressed-state picture is no longer sufficient. Instead, the electron and the quantized field must be considered jointly, opening the possibility that strong-field observables may exhibit new features and carry imprints of the light-matter entanglement established during the interaction.

In this Letter, we show that resonant multiphoton ionization driven by BSV enters precisely such a regime. Using a full quantum electrodynamical (QED) treatment, in which both the electron and the driving field are treated quantum mechanically, we show that the strong-field response is governed by the broad photon-number fluctuations of the BSV field and the resulting light-atom entanglement that develops through Rabi evolution. This gives rise to phenomena with no counterpart under coherent driving. Most strikingly, whereas coherent driving produces the usual ATI-order-independent Autler-Townes (AT) splitting in the photoelectron spectra, BSV driving yields an AT splitting that increases with ATI order. We show that this behavior follows a characteristic $\sqrt{2p}$ scaling law with ATI order $p$, providing an observable signature of the quantum nature of the BSV driving field in the strong-field regime.

Within the QED framework, the joint atom-field wavefunction is written as $|\Psi (t)\rangle =\sum_{n=0}^{N_{\max}}{\int{d}}x\,\psi _n(x,t)\,|x\rangle \otimes |n\rangle $, with $|x\rangle$ and $|n\rangle$ the electronic position and photon-number states, respectively. In the dipole approximation, the Hamiltonian (in atomic units) yields
\begin{align}
H=\frac{p_{x}^{2}}{2}+V_x(x)+\omega \left( a^{\dagger}a+\frac{1}{2} \right) +\lambda f(t)\,x\,(a+a^{\dagger})
	\label{Hamiltonian},
\end{align}
where $\lambda$ denotes the single-photon coupling strength \cite{Stammer2023QEDIntenseLM}. We model the target atom by a one-dimensional soft-core potential $V_x(x)=-1/\sqrt{x^2+1}$ yielding $E_g=-0.6698$ a.u. and $E_e=-0.2749$ a.u. for the ground and first excited states, and consider either BSV or coherent pulses resonant with the transition ($\omega=0.3949$ a.u.) with 31 fs FWHM duration.
To make the simulations tractable for large photon numbers and strong squeezing, we transform the field to the $y$-representation \cite{delaPena2025QEDHHG} and propagate in the interaction picture (see Supplementary Material (SM) Sec.\;S1 for details \cite{SM}), leading to 
$H_I(t)=\frac{p_{x}^{2}}{2}+V_x(x)+x\,\lambda f(t)\sqrt{2\omega}\!\left[ y\cos\mathrm{(}\omega t)+\frac{p_y}{\omega}\sin\mathrm{(}\omega t) \right]$.
This approach enables fully quantum simulations of strong-field ionization in the relevant parameter regime.

\begin{figure}[t]
	\centering
	\includegraphics[width=1.0\linewidth]{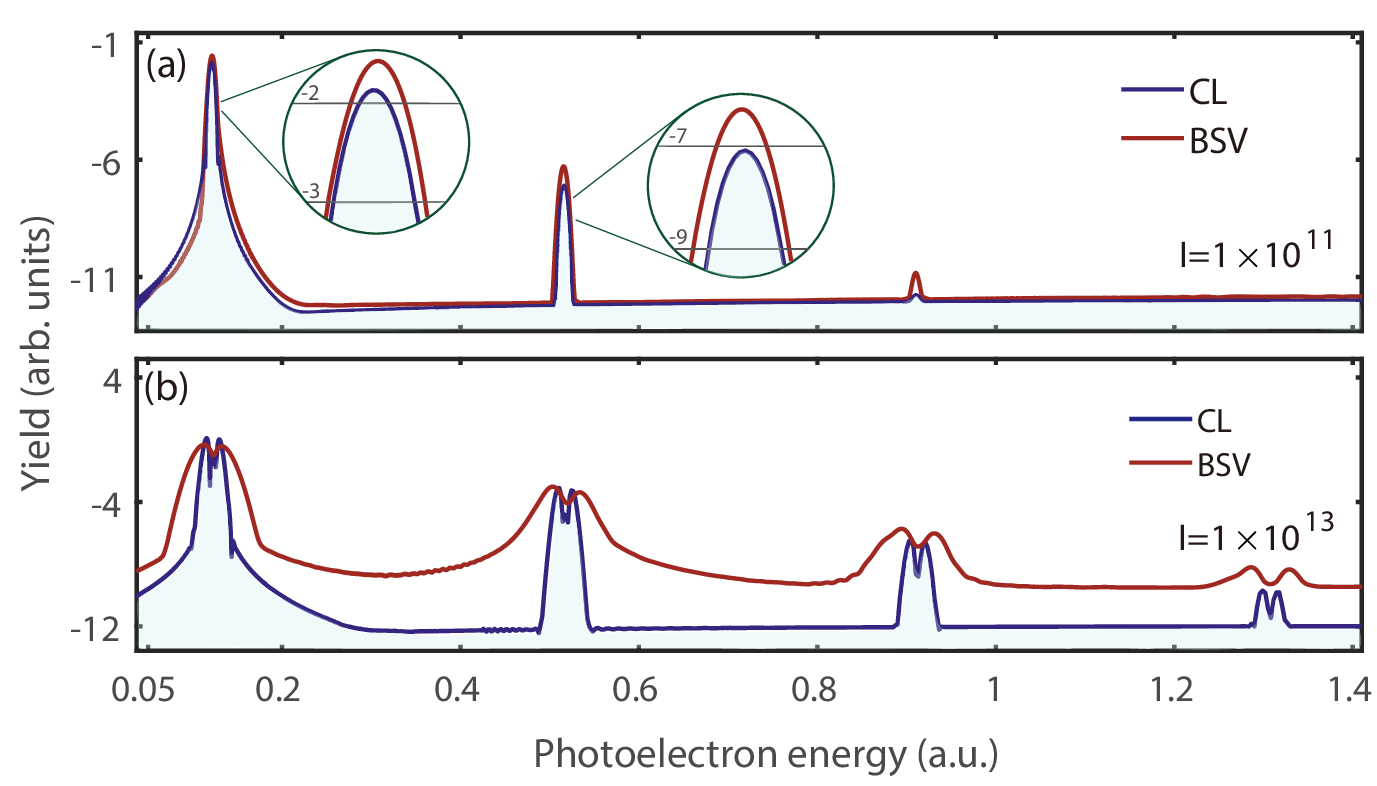}
	\caption{ Photoelectron energy spectra from QED simulations at intensities of (a) $1 \times 10^{11}\;\mathrm{W}/\mathrm{cm}^2$ and (b) $1 \times 10^{13}\;\mathrm{W}/\mathrm{cm}^2$.
		Red and blue curves correspond to BSV and intense coherent driving fields, respectively.
        The insets in panel (a) show magnified views of the first and second ATI peaks.}
	\label{Fig2}		
\end{figure}

The calculated photoelectron spectra for resonant multiphoton ionization driven by BSV and coherent pulses are shown in Fig.~\ref{Fig2}, for laser intensities of $1.0\times10^{11}\ \mathrm{W/cm^2}$ [Fig.~\ref{Fig2}(a)] and $1.0\times10^{13}\ \mathrm{W/cm^2}$ [Fig.~\ref{Fig2}(b)].
At lower intensities, three well-resolved ATI peaks are visible. The BSV-driven yields are substantially larger than those under coherent driving, and the enhancement increases with ATI order. This behavior is due to the strong photon-number fluctuations of the BSV, in which the nonlinear processes enhanced the statistical contributions of the BSV shots with large photon numbers~\cite{Spasibko2017Multiphoton,Manceau2019IndefiniteMean}. At higher intensities, each ATI peak develops a clear AT doublet, signaling the onset of Rabi oscillations. Two striking differences then distinguish BSV from coherent driving. First, for coherent driving, the AT splitting remains nearly the same across different ATI orders, whereas for BSV driving it increases monotonically with ATI order. Second, the BSV-induced enhancement of the ionization yield is strongly reduced once Rabi oscillations set in. For example, the first ATI peak is substantially enhanced by BSV in Fig.~\ref{Fig2}(a), but becomes nearly indistinguishable from the coherent-driving result in Fig.~\ref{Fig2}(b).

\begin{figure}[t]
	\centering
    	\includegraphics[width=1.0\linewidth]{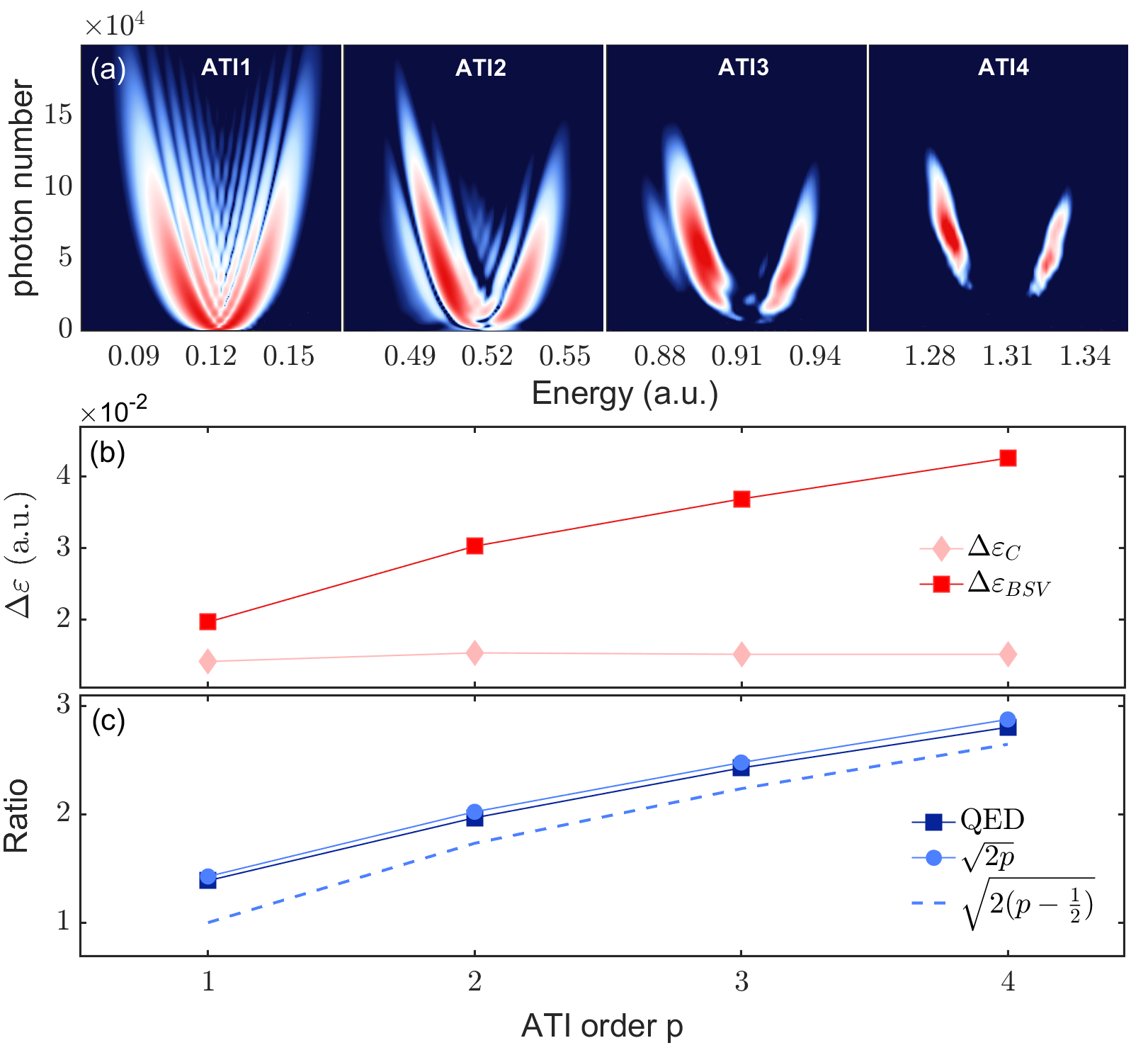}
	\caption{ (a) Field-photoelectron coincidence spectrum obtained from QED simulations under BSV driving.
		(b) Energy separation of the AT doublets as a function of ATI order for coherent (diamonds) and BSV (squares) driving.
		(c) Ratio of the AT-doublet energy separations under BSV and coherent driving,
		$\Delta\varepsilon_{\mathrm{BSV}}/\Delta\varepsilon_{\mathrm{C}}$, as a function of ATI order.
		QED results (squares) are compared with the analytical $\sqrt{2p}$ (circles) and $\sqrt{2p-1}$ (dashed line) scaling laws.}
\label{Fig3}		
\end{figure}

The ATI-order-independent AT splitting in the coherent driving is naturally understood within the standard semiclassical picture: the field acts as an essentially unchanged classical background, producing a single pair of dressed bound states, and the AT doublets in different ATI channels are simply replicas of the same underlying Rabi splitting \cite{Knight1980,Nandi2022,deLasHeras2025}. Under BSV driving, however, the broad photon-number distribution implies that different Fock components experience different effective couplings. The photoelectron must therefore be described jointly with the quantized field, 
$\left| \Psi_{el-ph} \right\rangle = \sum_{p,n} c_{p}^{(n)} \left| \psi_{p}^{(n)} \right\rangle \otimes \left| n \right\rangle$,
where $\left| n \right\rangle$ denotes the photon-number state of the driving field, and $\left| \psi_{p}^{(n)} \right\rangle$ is the corresponding photoelectron continuum component for the $p$th-order ATI. In the resonant strong-coupling regime, each photon-number sector $n$ carries a distinct effective Rabi frequency, so that $\left| \psi_{p}^{(n)} \right\rangle$ splits into two continuum branches. The $p$th ATI channel therefore takes the form
\begin{equation}
\left| \Psi_{el-ph}^{(p)} \right\rangle = \sum_n c_{p}^{(n)} \left(\left| \varepsilon_{p,-}^{(n)} \right\rangle - \left| \varepsilon_{p,+}^{(n)} \right\rangle\right) \otimes \left| n \right\rangle
\end{equation}
This picture is directly shown by the electron-field coincidence spectrum extracted from the QED simulations [Fig.~\ref{Fig3}(a)]. For a fixed ATI order, the distribution exhibits a clear two-branch structure over a broad photon-number range, with the branch separation increasing monotonically with $n$, and thus the joint state is nonseparable. The photon-number-resolved double-branch structure in the joint distribution therefore provides a spectroscopic signature of the underlying electron-field entanglement under BSV driving.

A comparison across ATI orders further reveals that different ATI orders sample different photon-number windows, and the dominant window shifts systematically to larger $n$ as the ATI order increases. This shift reflects ATI-selective sampling of photon-number fluctuations by the nonlinear transitions. The ATI-order-dependent AT splitting under BSV driving thus originates from a {\it cooperative} interplay between two effects: the entanglement-encoded, photon-number-resolved AT structure and the ATI-selective sampling of different photon-number sectors. We note that the electron-field entanglement here differs from that in the low-intensity regime described in quantum optics textbooks, where Rabi oscillations driven by a coherent field can generate entanglement \cite{ScullyZubairy1997}. In the high-intensity regime, coherent driving leaves the field nearly unchanged, so the joint electron-field state remains approximately separable even when Rabi oscillations occur. Under BSV driving, however, the broad photon-number distribution induces photon-number-dependent dynamics and makes the joint state nonseparable (as shown in Fig.~\ref{Fig1}).

We construct an extended Jaynes-Cummings (JC) model that explicitly includes the ionization continua associated with different ATI channels to quantitatively describe the picture above. In this model, the joint matter-light wavefunction is written as
\begin{align}
	|\Psi (t)\rangle
	&=
	\sum_{n=0}^{N_{\max}}
	\Big[
	c_{g}^{(n)}(t)\,|g,n\rangle
	+
	c_{e}^{(n)}(t)\,|e,n\rangle
	\nonumber\\
	&\quad
	+
	\sum_{p=1}^4
	\int d\varepsilon_{p}\,
	c_{f_p}^{(n)}(\varepsilon_{p},t)\,
	|\varepsilon_{p},n\rangle
	\Big] ,
	\label{modelphi}
\end{align}
where $|g,n\rangle$ and $|e,n\rangle$ denote the joint ground- and excited-state atom-field configurations, and $|\varepsilon_{p},n\rangle$ the continuum state in the $p$th ATI channel. This reduced model, which retains only the essential states, quantitatively reproduces the full QED results, confirming that it captures the underlying physics (see SM Sec.~S2 \cite{SM}).
Within this framework, the photon-number-resolved spectrum for the $p$th ATI order can be written analytically as \cite{SM}
\begin{align}
	P_p(n,\varepsilon_p )=A_{np}\cdot
	(D_{ef_p}\lambda^p)^2\left| \int_0^T{e^{-i(\varepsilon_0 -\varepsilon _{p})t}}\sin\left[\frac{1}{2}\Omega _{np}t\right]dt \right|^2
	\label{modelPkp},
\end{align}
where $\varepsilon_0=p\omega+E_e$ is the central energy of the $p$th ATI peak and
$\Omega _{np}=2D_{ge}\,\lambda\,\sqrt{n+1+p}$
is the photon-number-dependent Rabi frequency, with $D_{ij}$ ($i,j\in\{g,e,f_p\}$) denoting the atomic dipole matrix elements. Two distinct ingredients are encoded in Eq.~(\ref{modelPkp}). First, the prefactor, $A_{np}=|\alpha _{n+1+p}|^2\prod_{j=1}^p{(}n+j)$ (with $\alpha_n$ denoting the Fock-state expansion coefficient of the initial field state), determines how the $p$th ATI channel selectively samples the photon-number distribution. Second, the sine term generates the $n$-resolved AT splitting associated with the photon-number-dependent Rabi dynamics.

\begin{figure}[t]
	\centering
	\includegraphics[width=1.08\linewidth]{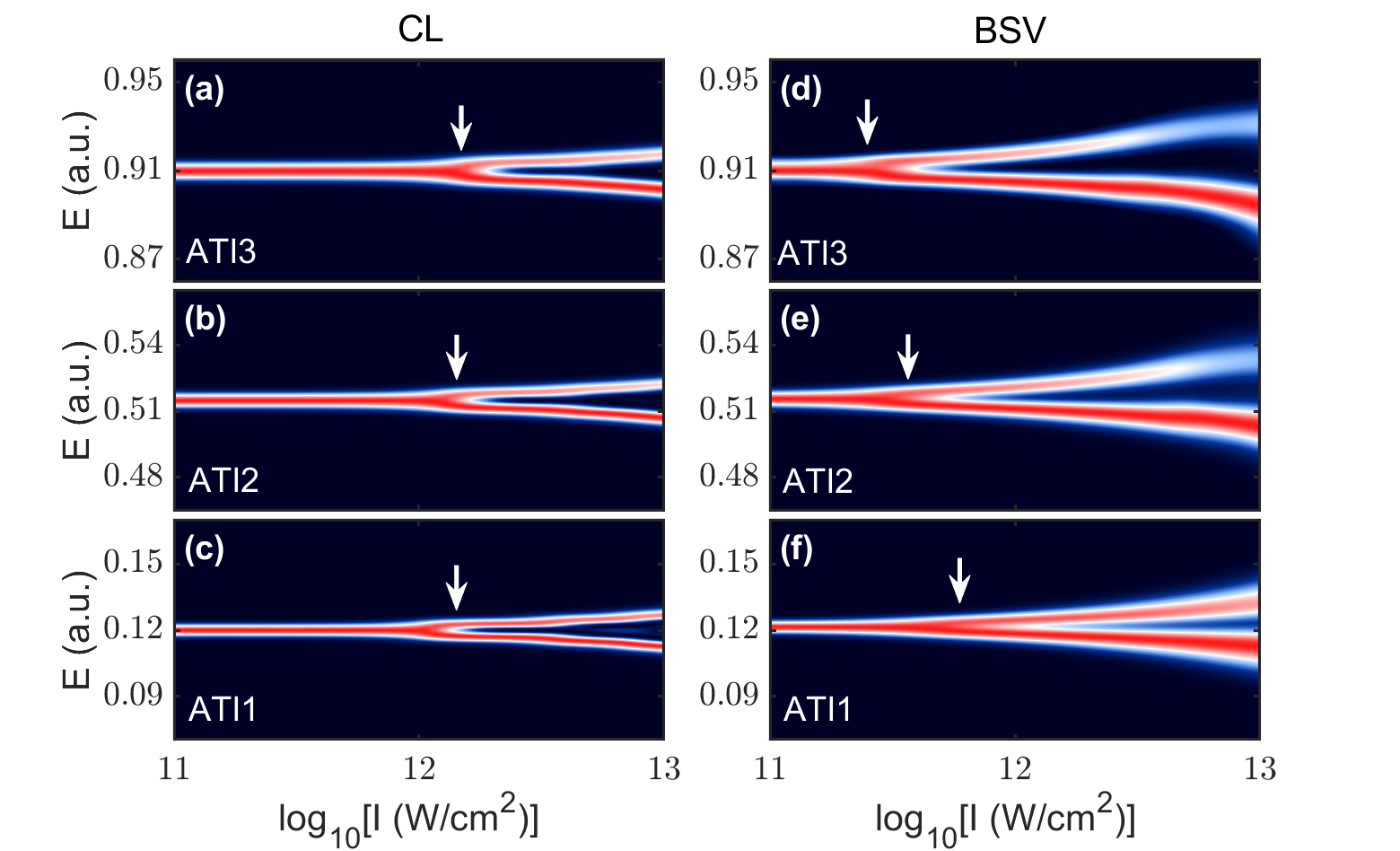}
	\caption{ Normalized photoelectron energy spectra as a function of laser intensity under (left column) coherent and (right column) BSV drivings. White arrows indicate the intensity threshold where resolvable Autler-Townes splitting emerges; the shift to lower intensities for higher ATI orders in BSV driving indicates a faster buildup of Rabi oscillations.}
	\label{Fig4}		
\end{figure}

Tracing over the field degrees of freedom, in the long-pulse limit we obtain the photoelectron spectrum \cite{SM},
\begin{align}
	S_p\propto \sum_{n=0}^{N_{\max}}{A_{np}}[\delta (\varepsilon _{p}-\varepsilon _0-\frac{1}{2}\Omega _{np})+\delta (\varepsilon _{p}-\varepsilon _0+\frac{1}{2}\Omega _{np})].
	\label{modelSkp}
\end{align}
By analyzing the maximum of the spectrum of Eq.\;\eqref{modelSkp}, we find that the splitting of the $p$th ATI channel in the BSV field is determined by the photon-number sector centered at (see SM Sec.\;S3 \cite{SM}) 
\begin{equation}
	N_p^* = 2p\bar n,
	\label{modelNp} \end{equation}
    where $\bar n$ is the mean photon number of the driving field.
The corresponding AT splitting is therefore
\begin{align}
	\Delta\varepsilon_{\mathrm{BSV}}(p) 
	= \Omega_{np}(N_p^*) 
	=  2D_{ge}\lambda\sqrt{2p\bar n}
	\label{modeldE}.
\end{align}
Compared to the AT splitting for intense coherent driving $\Delta\varepsilon_{\mathrm{C}}=2D_{ge}\lambda\sqrt{\bar n}$, this yields the ratio of $\Delta \varepsilon_{\mathrm{BSV}}/\Delta \varepsilon_{\mathrm{C}}=\sqrt{2p}$.

Figure~\ref{Fig3}(b) qualitatively compares the AT splittings as a function of ATI order in the coherent and the BSV fields. The ratio of the splitting $\Delta\varepsilon_{\mathrm{BSV}}/\Delta\varepsilon_{\mathrm{C}}$ exhibits a clear sublinear growth with ATI order, as shown in Fig.~\ref{Fig3}(c), and the analytical $\sqrt{2p}$ scaling law is in very good agreement with the full QED results. 
It is important to clarify that considering only ATI-order-selective sampling of photon-number states is insufficient. Indeed, if one analyzes the maximum of $A_{np}$ in Eq.~\ref{modelPkp}, the dominant photon number is instead found at $N_p^*=2(p-1/2)\bar{n}$, leading to a $\sqrt{2(p-1/2)}$ scaling law \cite{SM}. As shown by the dashed line in Fig.~\ref{Fig3}(c), this prediction deviates significantly from the full QED results.

\begin{figure}[t]
	\centering
	\includegraphics[width=0.75\linewidth]{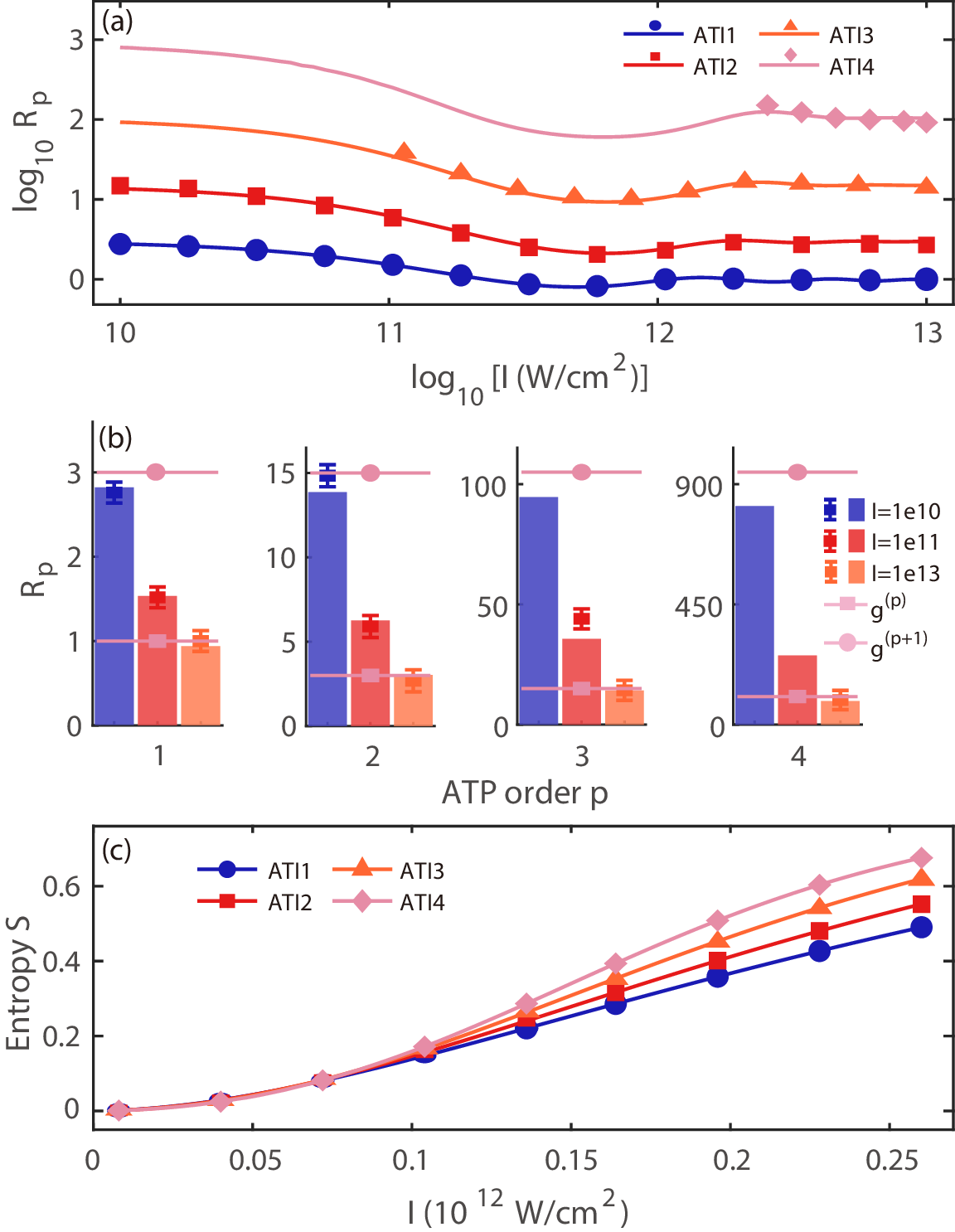}
	\caption{ (a) Yield ratio $R_p(I)$ between the BSV and coherent drivings, $R_p(I)$, as a function of the laser intensity for different ATI orders. Solid lines: extended JC model; symbols: QED calculations. (b) $R_p(I)$ at three representative intensities, $I=1\times10^{10}$, $1\times10^{11}$, and $1\times10^{13}$ W/cm$^2$. Bars: extended JC model; colored squares: QED calculations. (c) Von Neumann entanglement entropy between the BSV driving field and the $p$th ATI photoelectron as a function of laser intensity. The intensity is restricted to the range where $R_p$ drops rapidly in (a).
	}
	\label{Fig5}		
\end{figure}

The ATI-order dependent coupling strength in BSV is also revealed in the intensity-dependent buildup of Rabi oscillations in each ATI channel, as shown in Fig.~\ref{Fig4}. Under coherent driving, resolvable AT doublets appear at nearly the same intensity for different ATI orders [white arrows in Figs.~\ref{Fig4}(a)-~\ref{Fig4}(c)], consistent with a single effective Rabi frequency governing all channels. Under BSV driving, by contrast, the onset of resolvable AT splitting shifts systematically to lower intensity with increasing ATI order [white arrows in Figs.~\ref{Fig4}(d)-~\ref{Fig4}(f)], showing that higher-order ATI channels experience a faster buildup of Rabi oscillations.

Beyond reshaping the spectral structure, the buildup of Rabi oscillations also changes how strongly BSV enhances ionization yields relative to coherent driving. 
To quantify this effect, we define the yield ratio for the $p$th ATI channel, $R_p(I)\equiv \frac{Y_{p,\mathrm{BSV}}(I)}{Y_{p,\mathrm{C}}(I)}$, shown in Fig.~\ref{Fig5}(a), where $Y_{p,\mathrm{BSV}}(I)$ and $Y_{p,\mathrm{C}}(I)$ denote the ionization yields in the $p$th ATI channel under BSV and coherent driving, respectively. For each ATI order, $R_p(I)$ is not constant: it decreases rapidly with intensity before saturating at a high-intensity plateau.
The BSV enhancement is therefore not a fixed multiplicative factor set solely by the incident photon statistics, but evolves with the strong-field dynamics as Rabi oscillations develop.
This evolution is sampled and presented in Fig.~\ref{Fig5}(b), which shows $R_p$ at three representative intensities. 
In the low-intensity regime ($1\times10^{10}$ W/cm$^2$, blue), $R_p$  is close to the $(p+1)$th-order correlation function  $g^{(p+1)}$ of the BSV field. In the high-intensity regime ($1\times10^{13}$ W/cm$^2$, orange), it instead approaches $g^{(p)}$, with intermediate values appearing during the crossover.

The origin of this behavior is a change in the dominant ionization pathways during the buildup of Rabi oscillations. 
Before Rabi oscillations are established, ionization proceeds predominantly through direct ionization from the ground state, i.e., a $(p+1)$-photon process. Therefore, the ionization rate is proportional to the $(p+1)$th-order correlation function of the fields (see SM Sec.\;S4 \cite{SM}),
\begin{align}
	Y_{\mathrm{p}}^{\mathrm{D}} \propto g^{(p+1)}\langle n\rangle ^{p+1}
\end{align}
Accordingly, the ratio of the photoelectron yield between BSV and the coherent fields is governed by $g^{(p+1)}$.
Once Rabi oscillations are established, the resonant channel of Eq.~(\ref{modelPkp}) dominates instead, giving
\begin{align}
	Y_{\mathrm{p}}^{\mathrm{R}} \propto g^{(p)}\langle n\rangle ^{p}.
\end{align}
As a result, the ratio approaches $g^{(p)}$.
The monotonic decrease of $R_p(I)$ in Fig.~\ref{Fig5}(a) therefore reflects a continuous transfer from direct ground-state ionization to resonant ionization, driving the effective BSV enhancement from $g^{(p+1)}$ to $g^{(p)}$. Note that this transition itself is ATI-order dependent. Under BSV driving, Rabi oscillations build up faster for higher ATI orders (as shown in Fig.~\ref{Fig4}). This behavior is also reflected in the associated electron-field entanglement entropy [Fig.~\ref{Fig5}(c)], which increases more rapidly for higher ATI orders. Consistently, the corresponding $R_p(I)$ curves exhibit a faster crossover from the direct-ionization limit, governed by $g^{(p+1)}$, to the resonant limit, governed by $g^{(p)}$.

In summary, we have investigated resonant multiphoton ionization driven by BSV using full QED simulations. We reveal a striking $\sqrt{2p}$ scaling law of the Autler-Townes splitting with ATI order in the photoelectron energy spectra, in stark contrast to the conventional classical picture in which a single effective Rabi frequency governs all ATI channels. This feature carries a direct imprint of light-matter entanglement induced by the giant photon-number fluctuations of the BSV field. Moreover, we show that the buildup of Rabi oscillations modifies the BSV enhancement of nonlinear photoionization yields, driving the enhancement factor from $g^{(p+1)}$ to $g^{(p)}$ with increasing intensity. These findings are timely in light of recent experimental progress in the generation and single-shot characterization of femtosecond BSV pulses \cite{Kern2026Optica}, which brings BSV-based strong-field and attosecond experiments closer to experimental feasibility. Our work highlights the significance of light-matter entanglement in the strong-field regime. Extending these concepts to multielectron systems or to other strong-field phenomena, such as HHG, may open new avenues for quantum control and for the synthesis of macroscopic entanglement resources, advancing the interface between strong-field physics and quantum information science.

\section*{acknowledgments}

This work was supported by the National Natural Science Foundation of China (Grants No. U25D8005, 12374264, 12574378, 12204545, 12434010), National Key Research and Development Program of China (Grant No. 2023YFA1406800), Basic Research Support Program of Huazhong University of Science and Technology (2024BRA002). M.~F.~C.~acknowledges support by the National Key Research and Development Program of China (Grant No.~2023YFA1407100), Guangdong Province Science and Technology Major Project (Future functional materials under extreme conditions - 2021B0301030005), National Natural Science Foundation of China (Grant No. 12574092) and Guangdong Provincial Quantum Science Strategic Initiative (No. GDZX2504001). The computing work in this paper is supported by the Public Service Platform of High Performance Computing provided by Network and Computing Center of HUST.

X. Z., L. L., and Y. D. contribute equally.

	
		%
	

\begin{thebibliography}{71}%
		\makeatletter
		\providecommand \@ifxundefined [1]{%
			\@ifx{#1\undefined}
		}%
		\providecommand \@ifnum [1]{%
			\ifnum #1\expandafter \@firstoftwo
			\else \expandafter \@secondoftwo
			\fi
		}%
		\providecommand \@ifx [1]{%
			\ifx #1\expandafter \@firstoftwo
			\else \expandafter \@secondoftwo
			\fi
		}%
		\providecommand \natexlab [1]{#1}%
		\providecommand \enquote  [1]{``#1''}%
		\providecommand \bibnamefont  [1]{#1}%
		\providecommand \bibfnamefont [1]{#1}%
		\providecommand \citenamefont [1]{#1}%
		\providecommand \href@noop [0]{\@secondoftwo}%
		\providecommand \href [0]{\begingroup \@sanitize@url \@href}%
		\providecommand \@href[1]{\@@startlink{#1}\@@href}%
		\providecommand \@@href[1]{\endgroup#1\@@endlink}%
		\providecommand \@sanitize@url [0]{\catcode `\\12\catcode `\$12\catcode
			`\&12\catcode `\#12\catcode `\^12\catcode `\_12\catcode `\%12\relax}%
		\providecommand \@@startlink[1]{}%
		\providecommand \@@endlink[0]{}%
		\providecommand \url  [0]{\begingroup\@sanitize@url \@url }%
		\providecommand \@url [1]{\endgroup\@href {#1}{\urlprefix }}%
		\providecommand \urlprefix  [0]{URL }%
		\providecommand \Eprint [0]{\href }%
		\providecommand \doibase [0]{http://dx.doi.org/}%
		\providecommand \selectlanguage [0]{\@gobble}%
		\providecommand \bibinfo  [0]{\@secondoftwo}%
		\providecommand \bibfield  [0]{\@secondoftwo}%
		\providecommand \translation [1]{[#1]}%
		\providecommand \BibitemOpen [0]{}%
		\providecommand \bibitemStop [0]{}%
		\providecommand \bibitemNoStop [0]{.\EOS\space}%
		\providecommand \EOS [0]{\spacefactor3000\relax}%
		\providecommand \BibitemShut  [1]{\csname bibitem#1\endcsname}%
		\let\auto@bib@innerbib\@empty
		\bibitem [{\citenamefont {Paul}\ \emph {et~al.}(2001)\citenamefont {Paul},
			\citenamefont {Toma}, \citenamefont {Breger}, \citenamefont {Mullot},
			\citenamefont {Auge}, \citenamefont {Balcou}, \citenamefont {Muller},\ and\
			\citenamefont {Agostini}}]{Paul2001}%
		\BibitemOpen
		\bibfield  {author} {\bibinfo {author} {\bibfnamefont {Pierre-Mary}\
				\bibnamefont {Paul}}, \bibinfo {author} {\bibfnamefont {E}~\bibnamefont
				{Toma}}, \bibinfo {author} {\bibfnamefont {P}~\bibnamefont {Breger}},
			\bibinfo {author} {\bibfnamefont {Genevive}\ \bibnamefont {Mullot}}, \bibinfo
			{author} {\bibfnamefont {F}~\bibnamefont {Auge}}, \bibinfo {author}
			{\bibfnamefont {Ph}~\bibnamefont {Balcou}}, \bibinfo {author} {\bibfnamefont
				{H}~\bibnamefont {Muller}}, \ and\ \bibinfo {author} {\bibfnamefont {Pierre}\
				\bibnamefont {Agostini}},\ }\bibfield  {title} {\enquote {\bibinfo {title}
				{Observation of a train of attosecond pulses from high harmonic
					generation},}\ }\href {\doibase 10.1126/science.1059413} {\bibfield
			{journal} {\bibinfo  {journal} {Science (New York, N.Y.)}\ }\textbf {\bibinfo
				{volume} {292}},\ \bibinfo {pages} {1689--92} (\bibinfo {year}
			{2001})}\BibitemShut {NoStop}%
		\bibitem [{\citenamefont {Agostini}\ and\ \citenamefont
			{DiMauro}(2004)}]{Agostini2004AttosecondLightPulses}%
		\BibitemOpen
		\bibfield  {author} {\bibinfo {author} {\bibfnamefont {Pierre}\ \bibnamefont
				{Agostini}}\ and\ \bibinfo {author} {\bibfnamefont {Louis~F.}\ \bibnamefont
				{DiMauro}},\ }\bibfield  {title} {\enquote {\bibinfo {title} {The physics of
					attosecond light pulses},}\ }\href {\doibase 10.1088/0034-4885/67/6/R01}
		{\bibfield  {journal} {\bibinfo  {journal} {Reports on Progress in Physics}\
			}\textbf {\bibinfo {volume} {67}},\ \bibinfo {pages} {813--855} (\bibinfo
			{year} {2004})}\BibitemShut {NoStop}%
		\bibitem [{\citenamefont {Lambropoulos}(1976)}]{Bates1976}%
		\BibitemOpen
		\bibfield  {author} {\bibinfo {author} {\bibfnamefont {Peter}\ \bibnamefont
				{Lambropoulos}},\ }\bibfield  {title} {\enquote {\bibinfo {title} {Topics on
					multiphoton processes in atoms},}\ }in\ \href {\doibase
			10.1016/S0065-2199(08)60043-3} {\emph {\bibinfo {booktitle} {Advances in
					Atomic and Molecular Physics}}},\ \bibinfo {series} {Advances in Atomic and
			Molecular Physics}, Vol.~\bibinfo {volume} {12},\ \bibinfo {editor} {edited
			by\ \bibinfo {editor} {\bibfnamefont {D.R.}\ \bibnamefont {Bates}}\ and\
			\bibinfo {editor} {\bibfnamefont {Benjamin}\ \bibnamefont {Bederson}}}\
		(\bibinfo  {publisher} {Academic Press},\ \bibinfo {year} {1976})\ pp.\
		\bibinfo {pages} {87--164}\BibitemShut {NoStop}%
		\bibitem [{\citenamefont {Rhodes}(1985)}]{Charle1985}%
		\BibitemOpen
		\bibfield  {author} {\bibinfo {author} {\bibfnamefont {Charles~K}\
				\bibnamefont {Rhodes}},\ }\bibfield  {title} {\enquote {\bibinfo {title}
				{Multiphoton ionization of atoms},}\ }\href {\doibase
			10.1126/science.229.4720.1345} {\bibfield  {journal} {\bibinfo  {journal}
				{Science}\ }\textbf {\bibinfo {volume} {229}},\ \bibinfo {pages} {1345--1351}
			(\bibinfo {year} {1985})}\BibitemShut {NoStop}%
		\bibitem [{\citenamefont {Agostini}\ \emph {et~al.}(1979)\citenamefont
			{Agostini}, \citenamefont {Fabre}, \citenamefont {Mainfray}, \citenamefont
			{Petite},\ and\ \citenamefont {Rahman}}]{Agostini1979}%
		\BibitemOpen
		\bibfield  {author} {\bibinfo {author} {\bibfnamefont {Pierre}\ \bibnamefont
				{Agostini}}, \bibinfo {author} {\bibfnamefont {F}~\bibnamefont {Fabre}},
			\bibinfo {author} {\bibfnamefont {G{\'e}rard}\ \bibnamefont {Mainfray}},
			\bibinfo {author} {\bibfnamefont {Guillaume}\ \bibnamefont {Petite}}, \ and\
			\bibinfo {author} {\bibfnamefont {N~Ko}\ \bibnamefont {Rahman}},\ }\bibfield
		{title} {\enquote {\bibinfo {title} {Free-free transitions following
					six-photon ionization of xenon atoms},}\ }\href {\doibase
			10.1103/PhysRevLett.42.1127} {\bibfield  {journal} {\bibinfo  {journal}
				{Physical Review Letters}\ }\textbf {\bibinfo {volume} {42}},\ \bibinfo
			{pages} {1127--1130} (\bibinfo {year} {1979})}\BibitemShut {NoStop}%
		\bibitem [{\citenamefont {Freeman}\ \emph {et~al.}(1987)\citenamefont
			{Freeman}, \citenamefont {Bucksbaum}, \citenamefont {Milchberg},
			\citenamefont {Darack}, \citenamefont {Schumacher},\ and\ \citenamefont
			{Geusic}}]{Freeman1987}%
		\BibitemOpen
		\bibfield  {author} {\bibinfo {author} {\bibfnamefont {RR}~\bibnamefont
				{Freeman}}, \bibinfo {author} {\bibfnamefont {PH}~\bibnamefont {Bucksbaum}},
			\bibinfo {author} {\bibfnamefont {H}~\bibnamefont {Milchberg}}, \bibinfo
			{author} {\bibfnamefont {S}~\bibnamefont {Darack}}, \bibinfo {author}
			{\bibfnamefont {D}~\bibnamefont {Schumacher}}, \ and\ \bibinfo {author}
			{\bibfnamefont {ME}~\bibnamefont {Geusic}},\ }\bibfield  {title} {\enquote
			{\bibinfo {title} {Above-threshold ionization with subpicosecond laser
					pulses},}\ }\href {\doibase 10.1103/PhysRevLett.59.1092} {\bibfield
			{journal} {\bibinfo  {journal} {Physical Review Letters}\ }\textbf {\bibinfo
				{volume} {59}},\ \bibinfo {pages} {1092--1095} (\bibinfo {year}
			{1987})}\BibitemShut {NoStop}%
		\bibitem [{\citenamefont {Tsatrafyllis}\ \emph {et~al.}(2017)\citenamefont
			{Tsatrafyllis}, \citenamefont {Kominis}, \citenamefont {Gonoskov},\ and\
			\citenamefont {Tzallas}}]{Tsatrafyllis2017High-order}%
		\BibitemOpen
		\bibfield  {author} {\bibinfo {author} {\bibfnamefont {N}~\bibnamefont
				{Tsatrafyllis}}, \bibinfo {author} {\bibfnamefont {IK}~\bibnamefont
				{Kominis}}, \bibinfo {author} {\bibfnamefont {IA}~\bibnamefont {Gonoskov}}, \
			and\ \bibinfo {author} {\bibfnamefont {P}~\bibnamefont {Tzallas}},\
		}\bibfield  {title} {\enquote {\bibinfo {title} {High-order harmonics
					measured by the photon statistics of the infrared driving-field exiting the
					atomic medium},}\ }\href {\doibase 10.1038/ncomms15170} {\bibfield  {journal}
			{\bibinfo  {journal} {Nature Communications}\ }\textbf {\bibinfo {volume}
				{8}},\ \bibinfo {pages} {15170} (\bibinfo {year} {2017})}\BibitemShut
		{NoStop}%
		\bibitem [{\citenamefont {Gombk{\"o}t{\"o}}\ \emph {et~al.}(2020)\citenamefont
			{Gombk{\"o}t{\"o}}, \citenamefont {Varr{\'o}}, \citenamefont {Mati},\ and\
			\citenamefont {F{\"o}ldi}}]{Gombkoto2020HHGQuantizedField}%
		\BibitemOpen
		\bibfield  {author} {\bibinfo {author} {\bibfnamefont {{\'A}kos}\
				\bibnamefont {Gombk{\"o}t{\"o}}}, \bibinfo {author} {\bibfnamefont
				{S{\'a}ndor}\ \bibnamefont {Varr{\'o}}}, \bibinfo {author} {\bibfnamefont
				{P{\'e}ter}\ \bibnamefont {Mati}}, \ and\ \bibinfo {author} {\bibfnamefont
				{P{\'e}ter}\ \bibnamefont {F{\"o}ldi}},\ }\bibfield  {title} {\enquote
			{\bibinfo {title} {High-order harmonic generation as induced by a quantized
					field: Phase-space picture},}\ }\href {\doibase 10.1103/PhysRevA.101.013418}
		{\bibfield  {journal} {\bibinfo  {journal} {Physical Review A}\ }\textbf
			{\bibinfo {volume} {101}},\ \bibinfo {pages} {013418} (\bibinfo {year}
			{2020})}\BibitemShut {NoStop}%
		\bibitem [{\citenamefont {Gorlach}\ \emph {et~al.}(2020)\citenamefont
			{Gorlach}, \citenamefont {Neufeld}, \citenamefont {Rivera}, \citenamefont
			{Cohen},\ and\ \citenamefont {Kaminer}}]{Gorlach2020QuantumOpticalHHG}%
		\BibitemOpen
		\bibfield  {author} {\bibinfo {author} {\bibfnamefont {Alexey}\ \bibnamefont
				{Gorlach}}, \bibinfo {author} {\bibfnamefont {Ofer}\ \bibnamefont {Neufeld}},
			\bibinfo {author} {\bibfnamefont {Nicholas}\ \bibnamefont {Rivera}}, \bibinfo
			{author} {\bibfnamefont {Oren}\ \bibnamefont {Cohen}}, \ and\ \bibinfo
			{author} {\bibfnamefont {Ido}\ \bibnamefont {Kaminer}},\ }\bibfield  {title}
		{\enquote {\bibinfo {title} {The quantum-optical nature of high harmonic
					generation},}\ }\href {\doibase 10.1038/s41467-020-18218-w} {\bibfield
			{journal} {\bibinfo  {journal} {Nature Communications}\ }\textbf {\bibinfo
				{volume} {11}},\ \bibinfo {pages} {4598} (\bibinfo {year}
			{2020})}\BibitemShut {NoStop}%
		\bibitem [{\citenamefont {F{\"o}ldi}\ \emph {et~al.}(2021)\citenamefont
			{F{\"o}ldi}, \citenamefont {Magashegyi}, \citenamefont {Gombk{\"o}t{\"o}},\
			and\ \citenamefont {Varr{\'o}}}]{Foldi2021DescribingHHG}%
		\BibitemOpen
		\bibfield  {author} {\bibinfo {author} {\bibfnamefont {P{\'e}ter}\
				\bibnamefont {F{\"o}ldi}}, \bibinfo {author} {\bibfnamefont {Istv{\'a}n}\
				\bibnamefont {Magashegyi}}, \bibinfo {author} {\bibfnamefont {{\'A}kos}\
				\bibnamefont {Gombk{\"o}t{\"o}}}, \ and\ \bibinfo {author} {\bibfnamefont
				{S{\'a}ndor}\ \bibnamefont {Varr{\'o}}},\ }\bibfield  {title} {\enquote
			{\bibinfo {title} {Describing high-order harmonic generation using quantum
					optical models},}\ }\href {\doibase 10.3390/photonics8070263} {\bibfield
			{journal} {\bibinfo  {journal} {Photonics}\ }\textbf {\bibinfo {volume}
				{8}},\ \bibinfo {pages} {263} (\bibinfo {year} {2021})}\BibitemShut {NoStop}%
		\bibitem [{\citenamefont {Lewenstein}\ \emph {et~al.}(2021)\citenamefont
			{Lewenstein}, \citenamefont {Ciappina}, \citenamefont {Pisanty},
			\citenamefont {Rivera-Dean}, \citenamefont {Stammer}, \citenamefont
			{Lamprou},\ and\ \citenamefont {Tzallas}}]{Lewenstein2021OpticalCatStates}%
		\BibitemOpen
		\bibfield  {author} {\bibinfo {author} {\bibfnamefont {Maciej}\ \bibnamefont
				{Lewenstein}}, \bibinfo {author} {\bibfnamefont {Marcelo~F}\ \bibnamefont
				{Ciappina}}, \bibinfo {author} {\bibfnamefont {Emilio}\ \bibnamefont
				{Pisanty}}, \bibinfo {author} {\bibfnamefont {Javier}\ \bibnamefont
				{Rivera-Dean}}, \bibinfo {author} {\bibfnamefont {Philipp}\ \bibnamefont
				{Stammer}}, \bibinfo {author} {\bibfnamefont {Th}~\bibnamefont {Lamprou}}, \
			and\ \bibinfo {author} {\bibfnamefont {Paraskevas}\ \bibnamefont {Tzallas}},\
		}\bibfield  {title} {\enquote {\bibinfo {title} {Generation of optical
					schrödinger cat states in intense laser–matter interactions},}\ }\href
		{\doibase 10.1038/s41567-021-01317-w} {\bibfield  {journal} {\bibinfo
				{journal} {Nature Physics}\ }\textbf {\bibinfo {volume} {17}},\ \bibinfo
			{pages} {1104--1108} (\bibinfo {year} {2021})}\BibitemShut {NoStop}%
		\bibitem [{\citenamefont {Rivera-Dean}\ \emph {et~al.}(2022)\citenamefont
			{Rivera-Dean}, \citenamefont {Stammer}, \citenamefont {Maxwell},
			\citenamefont {Lamprou}, \citenamefont {Tzallas}, \citenamefont
			{Lewenstein},\ and\ \citenamefont
			{Ciappina}}]{RiveraDean2022LightMatterEntanglement}%
		\BibitemOpen
		\bibfield  {author} {\bibinfo {author} {\bibfnamefont {Javier}\ \bibnamefont
				{Rivera-Dean}}, \bibinfo {author} {\bibfnamefont {Philipp}\ \bibnamefont
				{Stammer}}, \bibinfo {author} {\bibfnamefont {Andrew~S.}\ \bibnamefont
				{Maxwell}}, \bibinfo {author} {\bibfnamefont {Theocharis}\ \bibnamefont
				{Lamprou}}, \bibinfo {author} {\bibfnamefont {Paraskevas}\ \bibnamefont
				{Tzallas}}, \bibinfo {author} {\bibfnamefont {Maciej}\ \bibnamefont
				{Lewenstein}}, \ and\ \bibinfo {author} {\bibfnamefont {Marcelo~F.}\
				\bibnamefont {Ciappina}},\ }\bibfield  {title} {\enquote {\bibinfo {title}
				{Light-matter entanglement after above-threshold ionization processes in
					atoms},}\ }\href {\doibase 10.1103/PhysRevA.106.063705} {\bibfield  {journal}
			{\bibinfo  {journal} {Physical Review A}\ }\textbf {\bibinfo {volume}
				{106}},\ \bibinfo {pages} {063705} (\bibinfo {year} {2022})}\BibitemShut
		{NoStop}%
		\bibitem [{\citenamefont
			{Stammer}(2022)}]{Stammer2022EntanglementMeasurementHHG}%
		\BibitemOpen
		\bibfield  {author} {\bibinfo {author} {\bibfnamefont {Philipp}\ \bibnamefont
				{Stammer}},\ }\bibfield  {title} {\enquote {\bibinfo {title} {Theory of
					entanglement and measurement in high-order harmonic generation},}\ }\href
		{\doibase 10.1103/PhysRevA.106.L050402} {\bibfield  {journal} {\bibinfo
				{journal} {Physical Review A}\ }\textbf {\bibinfo {volume} {106}},\ \bibinfo
			{pages} {L050402} (\bibinfo {year} {2022})}\BibitemShut {NoStop}%
		\bibitem [{\citenamefont {Stammer}\ \emph {et~al.}(2022)\citenamefont
			{Stammer}, \citenamefont {Rivera-Dean}, \citenamefont {Lamprou},
			\citenamefont {Pisanty}, \citenamefont {Ciappina}, \citenamefont {Tzallas},\
			and\ \citenamefont {Lewenstein}}]{Stammer2022HighPhotonEntangled}%
		\BibitemOpen
		\bibfield  {author} {\bibinfo {author} {\bibfnamefont {Philipp}\ \bibnamefont
				{Stammer}}, \bibinfo {author} {\bibfnamefont {Javier}\ \bibnamefont
				{Rivera-Dean}}, \bibinfo {author} {\bibfnamefont {Theocharis}\ \bibnamefont
				{Lamprou}}, \bibinfo {author} {\bibfnamefont {Emilio}\ \bibnamefont
				{Pisanty}}, \bibinfo {author} {\bibfnamefont {Marcelo~F.}\ \bibnamefont
				{Ciappina}}, \bibinfo {author} {\bibfnamefont {Paraskevas}\ \bibnamefont
				{Tzallas}}, \ and\ \bibinfo {author} {\bibfnamefont {Maciej}\ \bibnamefont
				{Lewenstein}},\ }\bibfield  {title} {\enquote {\bibinfo {title} {High photon
					number entangled states and coherent state superposition from the extreme
					ultraviolet to the far infrared},}\ }\href {\doibase
			10.1103/PhysRevLett.128.123603} {\bibfield  {journal} {\bibinfo  {journal}
				{Physical Review Letters}\ }\textbf {\bibinfo {volume} {128}},\ \bibinfo
			{pages} {123603} (\bibinfo {year} {2022})}\BibitemShut {NoStop}%
		\bibitem [{\citenamefont {Bhattacharya}\ \emph {et~al.}(2023)\citenamefont
			{Bhattacharya}, \citenamefont {Lamprou}, \citenamefont {Maxwell},
			\citenamefont {Ordonez}, \citenamefont {Pisanty}, \citenamefont
			{Rivera-Dean}, \citenamefont {Stammer}, \citenamefont {Ciappina},
			\citenamefont {Lewenstein},\ and\ \citenamefont
			{Tzallas}}]{Bhattacharya2023StrongLaserFieldPhysics}%
		\BibitemOpen
		\bibfield  {author} {\bibinfo {author} {\bibfnamefont {Utso}\ \bibnamefont
				{Bhattacharya}}, \bibinfo {author} {\bibfnamefont {Th}~\bibnamefont
				{Lamprou}}, \bibinfo {author} {\bibfnamefont {Andrew~S}\ \bibnamefont
				{Maxwell}}, \bibinfo {author} {\bibfnamefont {Andres}\ \bibnamefont
				{Ordonez}}, \bibinfo {author} {\bibfnamefont {Emilio}\ \bibnamefont
				{Pisanty}}, \bibinfo {author} {\bibfnamefont {Javier}\ \bibnamefont
				{Rivera-Dean}}, \bibinfo {author} {\bibfnamefont {Philipp}\ \bibnamefont
				{Stammer}}, \bibinfo {author} {\bibfnamefont {Marcelo~F}\ \bibnamefont
				{Ciappina}}, \bibinfo {author} {\bibfnamefont {Maciej}\ \bibnamefont
				{Lewenstein}}, \ and\ \bibinfo {author} {\bibfnamefont {Paraskevas}\
				\bibnamefont {Tzallas}},\ }\bibfield  {title} {\enquote {\bibinfo {title}
				{Strong–laser–field physics, non–classical light states and quantum
					information science},}\ }\href {\doibase 10.1088/1361-6633/acea31} {\bibfield
			{journal} {\bibinfo  {journal} {Reports on Progress in Physics}\ }\textbf
			{\bibinfo {volume} {86}},\ \bibinfo {pages} {094401} (\bibinfo {year}
			{2023})}\BibitemShut {NoStop}%
		\bibitem [{\citenamefont {Cruz-Rodriguez}\ \emph {et~al.}(2024)\citenamefont
			{Cruz-Rodriguez}, \citenamefont {Dey}, \citenamefont {Freibert},\ and\
			\citenamefont {Stammer}}]{CruzRodriguez2024QuantumPhenomenaAttosecond}%
		\BibitemOpen
		\bibfield  {author} {\bibinfo {author} {\bibfnamefont {Lidice}\ \bibnamefont
				{Cruz-Rodriguez}}, \bibinfo {author} {\bibfnamefont {Diptesh}\ \bibnamefont
				{Dey}}, \bibinfo {author} {\bibfnamefont {Antonia}\ \bibnamefont {Freibert}},
			\ and\ \bibinfo {author} {\bibfnamefont {Philipp}\ \bibnamefont {Stammer}},\
		}\bibfield  {title} {\enquote {\bibinfo {title} {Quantum phenomena in
					attosecond science},}\ }\href {\doibase 10.1038/s42254-024-00769-2}
		{\bibfield  {journal} {\bibinfo  {journal} {Nature Reviews Physics}\ }\textbf
			{\bibinfo {volume} {6}},\ \bibinfo {pages} {691--704} (\bibinfo {year}
			{2024})}\BibitemShut {NoStop}%
		\bibitem [{\citenamefont {Hubenschmid}\ \emph {et~al.}(2024)\citenamefont
			{Hubenschmid}, \citenamefont {Guedes},\ and\ \citenamefont
			{Burkard}}]{Hubenschmid2024OpticalTimeDomainTomography}%
		\BibitemOpen
		\bibfield  {author} {\bibinfo {author} {\bibfnamefont {Emanuel}\ \bibnamefont
				{Hubenschmid}}, \bibinfo {author} {\bibfnamefont {Thiago~LM}\ \bibnamefont
				{Guedes}}, \ and\ \bibinfo {author} {\bibfnamefont {Guido}\ \bibnamefont
				{Burkard}},\ }\bibfield  {title} {\enquote {\bibinfo {title} {Optical
					time-domain quantum state tomography on a subcycle scale},}\ }\href {\doibase
			10.1103/PhysRevX.14.041032} {\bibfield  {journal} {\bibinfo  {journal}
				{Physical Review X}\ }\textbf {\bibinfo {volume} {14}},\ \bibinfo {pages}
			{041032} (\bibinfo {year} {2024})}\BibitemShut {NoStop}%
		\bibitem [{\citenamefont {Lange}\ \emph {et~al.}(2024)\citenamefont {Lange},
			\citenamefont {Hansen},\ and\ \citenamefont
			{Madsen}}]{Lange2024ElectronCorrelationNonclassicality}%
		\BibitemOpen
		\bibfield  {author} {\bibinfo {author} {\bibfnamefont {Christian~Saugbjerg}\
				\bibnamefont {Lange}}, \bibinfo {author} {\bibfnamefont {Thomas}\
				\bibnamefont {Hansen}}, \ and\ \bibinfo {author} {\bibfnamefont {Lars~Bojer}\
				\bibnamefont {Madsen}},\ }\bibfield  {title} {\enquote {\bibinfo {title}
				{Electron-correlation-induced nonclassicality of light from high-order
					harmonic generation},}\ }\href {\doibase 10.1103/PhysRevA.109.033110}
		{\bibfield  {journal} {\bibinfo  {journal} {Physical Review A}\ }\textbf
			{\bibinfo {volume} {109}},\ \bibinfo {pages} {033110} (\bibinfo {year}
			{2024})}\BibitemShut {NoStop}%
		\bibitem [{\citenamefont {Pizzi}\ \emph {et~al.}(2023)\citenamefont {Pizzi},
			\citenamefont {Gorlach}, \citenamefont {Rivera}, \citenamefont {Nunnenkamp},\
			and\ \citenamefont {Kaminer}}]{Pizzi2023LightEmissionManyBody}%
		\BibitemOpen
		\bibfield  {author} {\bibinfo {author} {\bibfnamefont {Andrea}\ \bibnamefont
				{Pizzi}}, \bibinfo {author} {\bibfnamefont {Alexey}\ \bibnamefont {Gorlach}},
			\bibinfo {author} {\bibfnamefont {Nicholas}\ \bibnamefont {Rivera}}, \bibinfo
			{author} {\bibfnamefont {Andreas}\ \bibnamefont {Nunnenkamp}}, \ and\
			\bibinfo {author} {\bibfnamefont {Ido}\ \bibnamefont {Kaminer}},\ }\bibfield
		{title} {\enquote {\bibinfo {title} {Light emission from strongly driven
					many-body systems},}\ }\href {\doibase 10.1038/s41567-022-01910-7} {\bibfield
			{journal} {\bibinfo  {journal} {Nature Physics}\ }\textbf {\bibinfo {volume}
				{19}},\ \bibinfo {pages} {551--561} (\bibinfo {year} {2023})}\BibitemShut
		{NoStop}%
		\bibitem [{\citenamefont {Theidel}\ \emph {et~al.}(2024)\citenamefont
			{Theidel}, \citenamefont {Cotte}, \citenamefont {Sondenheimer}, \citenamefont
			{Shiriaeva}, \citenamefont {Froidevaux}, \citenamefont {Severin},
			\citenamefont {Merdji-Larue}, \citenamefont {Mosel}, \citenamefont
			{Fr\"ohlich}, \citenamefont {Weber}, \citenamefont {Morgner}, \citenamefont
			{Kovacev}, \citenamefont {Biegert},\ and\ \citenamefont
			{Merdji}}]{Theidel2024EvidenceQuantumOpticalHHG}%
		\BibitemOpen
		\bibfield  {author} {\bibinfo {author} {\bibfnamefont {David}\ \bibnamefont
				{Theidel}}, \bibinfo {author} {\bibfnamefont {Viviane}\ \bibnamefont
				{Cotte}}, \bibinfo {author} {\bibfnamefont {Ren\'e}\ \bibnamefont
				{Sondenheimer}}, \bibinfo {author} {\bibfnamefont {Viktoriia}\ \bibnamefont
				{Shiriaeva}}, \bibinfo {author} {\bibfnamefont {Marie}\ \bibnamefont
				{Froidevaux}}, \bibinfo {author} {\bibfnamefont {Vladislav}\ \bibnamefont
				{Severin}}, \bibinfo {author} {\bibfnamefont {Adam}\ \bibnamefont
				{Merdji-Larue}}, \bibinfo {author} {\bibfnamefont {Philip}\ \bibnamefont
				{Mosel}}, \bibinfo {author} {\bibfnamefont {Sven}\ \bibnamefont
				{Fr\"ohlich}}, \bibinfo {author} {\bibfnamefont {Kim-Alessandro}\
				\bibnamefont {Weber}}, \bibinfo {author} {\bibfnamefont {Uwe}\ \bibnamefont
				{Morgner}}, \bibinfo {author} {\bibfnamefont {Milutin}\ \bibnamefont
				{Kovacev}}, \bibinfo {author} {\bibfnamefont {Jens}\ \bibnamefont {Biegert}},
			\ and\ \bibinfo {author} {\bibfnamefont {Hamed}\ \bibnamefont {Merdji}},\
		}\bibfield  {title} {\enquote {\bibinfo {title} {Evidence of the quantum
					optical nature of high-harmonic generation},}\ }\href {\doibase
			10.1103/PRXQuantum.5.040319} {\bibfield  {journal} {\bibinfo  {journal} {PRX
					Quantum}\ }\textbf {\bibinfo {volume} {5}},\ \bibinfo {pages} {040319}
			(\bibinfo {year} {2024})}\BibitemShut {NoStop}%
		\bibitem [{\citenamefont {Stammer}\ \emph {et~al.}(2024)\citenamefont
			{Stammer}, \citenamefont {Rivera-Dean}, \citenamefont {Maxwell},
			\citenamefont {Lamprou}, \citenamefont {Argüello-Luengo}, \citenamefont
			{Tzallas}, \citenamefont {Ciappina},\ and\ \citenamefont
			{Lewenstein}}]{Stammer2024EntanglementSqueezingHHG}%
		\BibitemOpen
		\bibfield  {author} {\bibinfo {author} {\bibfnamefont {Philipp}\ \bibnamefont
				{Stammer}}, \bibinfo {author} {\bibfnamefont {Javier}\ \bibnamefont
				{Rivera-Dean}}, \bibinfo {author} {\bibfnamefont {Andrew~S.}\ \bibnamefont
				{Maxwell}}, \bibinfo {author} {\bibfnamefont {Theocharis}\ \bibnamefont
				{Lamprou}}, \bibinfo {author} {\bibfnamefont {Javier}\ \bibnamefont
				{Argüello-Luengo}}, \bibinfo {author} {\bibfnamefont {Paraskevas}\
				\bibnamefont {Tzallas}}, \bibinfo {author} {\bibfnamefont {Marcelo~F.}\
				\bibnamefont {Ciappina}}, \ and\ \bibinfo {author} {\bibfnamefont {Maciej}\
				\bibnamefont {Lewenstein}},\ }\bibfield  {title} {\enquote {\bibinfo {title}
				{Entanglement and squeezing of the optical field modes in high harmonic
					generation},}\ }\href {\doibase 10.1103/PhysRevLett.132.143603} {\bibfield
			{journal} {\bibinfo  {journal} {Physical Review Letters}\ }\textbf {\bibinfo
				{volume} {132}},\ \bibinfo {pages} {143603} (\bibinfo {year}
			{2024})}\BibitemShut {NoStop}%
		\bibitem [{\citenamefont {Rivera-Dean}\ \emph {et~al.}(2024)\citenamefont
			{Rivera-Dean}, \citenamefont {Crispin}, \citenamefont {Stammer},
			\citenamefont {Lamprou}, \citenamefont {Pisanty}, \citenamefont {Kr\"uger},
			\citenamefont {Tzallas}, \citenamefont {Lewenstein},\ and\ \citenamefont
			{Ciappina}}]{RiveraDean2024SqueezedHHGExcitedAtoms}%
		\BibitemOpen
		\bibfield  {author} {\bibinfo {author} {\bibfnamefont {Javier}\ \bibnamefont
				{Rivera-Dean}}, \bibinfo {author} {\bibfnamefont {H.~B.}\ \bibnamefont
				{Crispin}}, \bibinfo {author} {\bibfnamefont {Philipp}\ \bibnamefont
				{Stammer}}, \bibinfo {author} {\bibfnamefont {Theocharis}\ \bibnamefont
				{Lamprou}}, \bibinfo {author} {\bibfnamefont {Emilio}\ \bibnamefont
				{Pisanty}}, \bibinfo {author} {\bibfnamefont {Martin}\ \bibnamefont
				{Kr\"uger}}, \bibinfo {author} {\bibfnamefont {Paraskevas}\ \bibnamefont
				{Tzallas}}, \bibinfo {author} {\bibfnamefont {Maciej}\ \bibnamefont
				{Lewenstein}}, \ and\ \bibinfo {author} {\bibfnamefont {Marcelo~F.}\
				\bibnamefont {Ciappina}},\ }\bibfield  {title} {\enquote {\bibinfo {title}
				{Squeezed states of light after high-order harmonic generation in excited
					atomic systems},}\ }\href {\doibase 10.1103/PhysRevA.110.063118} {\bibfield
			{journal} {\bibinfo  {journal} {Physical Review A}\ }\textbf {\bibinfo
				{volume} {110}},\ \bibinfo {pages} {063118} (\bibinfo {year}
			{2024})}\BibitemShut {NoStop}%
		\bibitem [{\citenamefont {Gothelf}\ \emph {et~al.}(2025)\citenamefont
			{Gothelf}, \citenamefont {Lange},\ and\ \citenamefont
			{Madsen}}]{Gothelf2025HHGCrystalQuantumLight}%
		\BibitemOpen
		\bibfield  {author} {\bibinfo {author} {\bibfnamefont {Rasmus~Vesterager}\
				\bibnamefont {Gothelf}}, \bibinfo {author} {\bibfnamefont
				{Christian~Saugbjerg}\ \bibnamefont {Lange}}, \ and\ \bibinfo {author}
			{\bibfnamefont {Lars~Bojer}\ \bibnamefont {Madsen}},\ }\bibfield  {title}
		{\enquote {\bibinfo {title} {High-order harmonic generation in a crystal
					driven by quantum light},}\ }\href {\doibase 10.1103/PhysRevA.111.063105}
		{\bibfield  {journal} {\bibinfo  {journal} {Physical Review A}\ }\textbf
			{\bibinfo {volume} {111}},\ \bibinfo {pages} {063105} (\bibinfo {year}
			{2025})}\BibitemShut {NoStop}%
		\bibitem [{\citenamefont {Wang}\ \emph
			{et~al.}(2025{\natexlab{a}})\citenamefont {Wang}, \citenamefont {Yang},
			\citenamefont {Cui}, \citenamefont {He}, \citenamefont {Ou}, \citenamefont
			{Jin}, \citenamefont {Liao}, \citenamefont {Lan},\ and\ \citenamefont
			{Lu}}]{Wang2025QuantumPathControlHHG}%
		\BibitemOpen
		\bibfield  {author} {\bibinfo {author} {\bibfnamefont {Feng}\ \bibnamefont
				{Wang}}, \bibinfo {author} {\bibfnamefont {Chunhui}\ \bibnamefont {Yang}},
			\bibinfo {author} {\bibfnamefont {Xinyi}\ \bibnamefont {Cui}}, \bibinfo
			{author} {\bibfnamefont {Lixin}\ \bibnamefont {He}}, \bibinfo {author}
			{\bibfnamefont {Tianxin}\ \bibnamefont {Ou}}, \bibinfo {author}
			{\bibfnamefont {Rui-Bo}\ \bibnamefont {Jin}}, \bibinfo {author}
			{\bibfnamefont {Qing}\ \bibnamefont {Liao}}, \bibinfo {author} {\bibfnamefont
				{Pengfei}\ \bibnamefont {Lan}}, \ and\ \bibinfo {author} {\bibfnamefont
				{Peixiang}\ \bibnamefont {Lu}},\ }\bibfield  {title} {\enquote {\bibinfo
				{title} {Quantum path control in high-order harmonic generation via squeezed
					light},}\ }\href {\doibase 10.1103/hph3-kzb2} {\bibfield  {journal} {\bibinfo
				{journal} {Phys. Rev. A}\ }\textbf {\bibinfo {volume} {112}},\ \bibinfo
			{pages} {043119} (\bibinfo {year} {2025}{\natexlab{a}})}\BibitemShut
		{NoStop}%
		\bibitem [{\citenamefont {Wang}\ and\ \citenamefont
			{Bian}(2025)}]{Wang2025HHGQuantumLightVonNeumann}%
		\BibitemOpen
		\bibfield  {author} {\bibinfo {author} {\bibfnamefont {Yi-Ben}\ \bibnamefont
				{Wang}}\ and\ \bibinfo {author} {\bibfnamefont {Xue-Bin}\ \bibnamefont
				{Bian}},\ }\bibfield  {title} {\enquote {\bibinfo {title} {High-order
					harmonic generation in quantum light by a generalized von neumann lattice
					method},}\ }\href {\doibase 10.1103/PhysRevA.111.043111} {\bibfield
			{journal} {\bibinfo  {journal} {Physical Review A}\ }\textbf {\bibinfo
				{volume} {111}},\ \bibinfo {pages} {043111} (\bibinfo {year}
			{2025})}\BibitemShut {NoStop}%
		\bibitem [{\citenamefont {Wang}\ \emph
			{et~al.}(2025{\natexlab{b}})\citenamefont {Wang}, \citenamefont {Lai},\ and\
			\citenamefont {Liu}}]{ShiJunWang2025}%
		\BibitemOpen
		\bibfield  {author} {\bibinfo {author} {\bibfnamefont {ShiJun}\ \bibnamefont
				{Wang}}, \bibinfo {author} {\bibfnamefont {XuanYang}\ \bibnamefont {Lai}}, \
			and\ \bibinfo {author} {\bibfnamefont {XiaoJun}\ \bibnamefont {Liu}},\
		}\bibfield  {title} {\enquote {\bibinfo {title} {Attosecond pulse synthesis
					from high-order harmonic generation in intense squeezed light},}\ }\href
		{\doibase 10.1103/4x11-phs2} {\bibfield  {journal} {\bibinfo  {journal}
				{Phys. Rev. A}\ }\textbf {\bibinfo {volume} {112}},\ \bibinfo {pages}
			{L011102} (\bibinfo {year} {2025}{\natexlab{b}})}\BibitemShut {NoStop}%
		\bibitem [{\citenamefont {Yi}\ \emph {et~al.}(2025)\citenamefont {Yi},
			\citenamefont {Klimkin}, \citenamefont {Brown}, \citenamefont {Smirnova},
			\citenamefont {Patchkovskii}, \citenamefont {Babushkin},\ and\ \citenamefont
			{Ivanov}}]{Yi2025MassivelyEntangledBrightStates}%
		\BibitemOpen
		\bibfield  {author} {\bibinfo {author} {\bibfnamefont {Sili}\ \bibnamefont
				{Yi}}, \bibinfo {author} {\bibfnamefont {Nikolai~D.}\ \bibnamefont
				{Klimkin}}, \bibinfo {author} {\bibfnamefont {Graham~Gardiner}\ \bibnamefont
				{Brown}}, \bibinfo {author} {\bibfnamefont {Olga}\ \bibnamefont {Smirnova}},
			\bibinfo {author} {\bibfnamefont {Serguei}\ \bibnamefont {Patchkovskii}},
			\bibinfo {author} {\bibfnamefont {Ihar}\ \bibnamefont {Babushkin}}, \ and\
			\bibinfo {author} {\bibfnamefont {Misha}\ \bibnamefont {Ivanov}},\ }\bibfield
		{title} {\enquote {\bibinfo {title} {Generation of massively entangled
					bright states of light during harmonic generation in resonant media},}\
		}\href {\doibase 10.1103/PhysRevX.15.011023} {\bibfield  {journal} {\bibinfo
				{journal} {Physical Review X}\ }\textbf {\bibinfo {volume} {15}},\ \bibinfo
			{pages} {011023} (\bibinfo {year} {2025})}\BibitemShut {NoStop}%
		\bibitem [{\citenamefont {Spasibko}\ \emph {et~al.}(2017)\citenamefont
			{Spasibko}, \citenamefont {Kopylov}, \citenamefont {Krutyanskiy},
			\citenamefont {Murzina}, \citenamefont {Leuchs},\ and\ \citenamefont
			{Chekhova}}]{Spasibko2017Multiphoton}%
		\BibitemOpen
		\bibfield  {author} {\bibinfo {author} {\bibfnamefont {Kirill~Yu.}\
				\bibnamefont {Spasibko}}, \bibinfo {author} {\bibfnamefont {Denis~A.}\
				\bibnamefont {Kopylov}}, \bibinfo {author} {\bibfnamefont {Victor~L.}\
				\bibnamefont {Krutyanskiy}}, \bibinfo {author} {\bibfnamefont {Tatiana~V.}\
				\bibnamefont {Murzina}}, \bibinfo {author} {\bibfnamefont {Gerd}\
				\bibnamefont {Leuchs}}, \ and\ \bibinfo {author} {\bibfnamefont {Maria~V.}\
				\bibnamefont {Chekhova}},\ }\bibfield  {title} {\enquote {\bibinfo {title}
				{Multiphoton effects enhanced due to ultrafast photon-number fluctuations},}\
		}\href {\doibase 10.1103/PhysRevLett.119.223603} {\bibfield  {journal}
			{\bibinfo  {journal} {Physical Review Letters}\ }\textbf {\bibinfo {volume}
				{119}},\ \bibinfo {pages} {223603} (\bibinfo {year} {2017})}\BibitemShut
		{NoStop}%
		\bibitem [{\citenamefont {Khalaf}\ and\ \citenamefont
			{Kaminer}(2023)}]{Khalaf2023ComptonQuantumLight}%
		\BibitemOpen
		\bibfield  {author} {\bibinfo {author} {\bibfnamefont {Majed}\ \bibnamefont
				{Khalaf}}\ and\ \bibinfo {author} {\bibfnamefont {Ido}\ \bibnamefont
				{Kaminer}},\ }\bibfield  {title} {\enquote {\bibinfo {title} {Compton
					scattering driven by intense quantum light},}\ }\href {\doibase
			10.1126/sciadv.ade0932} {\bibfield  {journal} {\bibinfo  {journal} {Science
					Advances}\ }\textbf {\bibinfo {volume} {9}},\ \bibinfo {pages} {eade0932}
			(\bibinfo {year} {2023})}\BibitemShut {NoStop}%
		\bibitem [{\citenamefont {Even~Tzur}\ \emph {et~al.}(2024)\citenamefont
			{Even~Tzur}, \citenamefont {Birk}, \citenamefont {Gorlach}, \citenamefont
			{Kaminer}, \citenamefont {Kr\"uger},\ and\ \citenamefont
			{Cohen}}]{EvenTzur2024SqueezedHighOrderHarmonics}%
		\BibitemOpen
		\bibfield  {author} {\bibinfo {author} {\bibfnamefont {Matan}\ \bibnamefont
				{Even~Tzur}}, \bibinfo {author} {\bibfnamefont {Michael}\ \bibnamefont
				{Birk}}, \bibinfo {author} {\bibfnamefont {Alexey}\ \bibnamefont {Gorlach}},
			\bibinfo {author} {\bibfnamefont {Ido}\ \bibnamefont {Kaminer}}, \bibinfo
			{author} {\bibfnamefont {Michael}\ \bibnamefont {Kr\"uger}}, \ and\ \bibinfo
			{author} {\bibfnamefont {Oren}\ \bibnamefont {Cohen}},\ }\bibfield  {title}
		{\enquote {\bibinfo {title} {Generation of squeezed high-order harmonics},}\
		}\href {\doibase 10.1103/PhysRevResearch.6.033079} {\bibfield  {journal}
			{\bibinfo  {journal} {Physical Review Research}\ }\textbf {\bibinfo {volume}
				{6}},\ \bibinfo {pages} {033079} (\bibinfo {year} {2024})}\BibitemShut
		{NoStop}%
		\bibitem [{\citenamefont {Long}\ \emph {et~al.}(2025)\citenamefont {Long},
			\citenamefont {Li},\ and\ \citenamefont
			{Liu}}]{Long2025HydrogenMolecularDissociation}%
		\BibitemOpen
		\bibfield  {author} {\bibinfo {author} {\bibfnamefont {Xiaoxiao}\
				\bibnamefont {Long}}, \bibinfo {author} {\bibfnamefont {Peizeng}\
				\bibnamefont {Li}}, \ and\ \bibinfo {author} {\bibfnamefont {Yunquan}\
				\bibnamefont {Liu}},\ }\bibfield  {title} {\enquote {\bibinfo {title}
				{Hydrogen molecular dissociation driven by quantum light},}\ }\href {\doibase
			10.1103/4g87-d9j3} {\bibfield  {journal} {\bibinfo  {journal} {Phys. Rev.
					Lett.}\ }\textbf {\bibinfo {volume} {135}},\ \bibinfo {pages} {153201}
			(\bibinfo {year} {2025})}\BibitemShut {NoStop}%
		\bibitem [{\citenamefont {Lemieux}\ \emph {et~al.}(2025)\citenamefont
			{Lemieux}, \citenamefont {Jalil}, \citenamefont {Purschke}, \citenamefont
			{Boroumand}, \citenamefont {Villeneuve}, \citenamefont {Naumov},
			\citenamefont {Brabec},\ and\ \citenamefont
			{Vampa}}]{Lemieux2025PhotonBunchingHHG}%
		\BibitemOpen
		\bibfield  {author} {\bibinfo {author} {\bibfnamefont {Samuel}\ \bibnamefont
				{Lemieux}}, \bibinfo {author} {\bibfnamefont {Sohail~A.}\ \bibnamefont
				{Jalil}}, \bibinfo {author} {\bibfnamefont {David~N.}\ \bibnamefont
				{Purschke}}, \bibinfo {author} {\bibfnamefont {Neda}\ \bibnamefont
				{Boroumand}}, \bibinfo {author} {\bibfnamefont {David}\ \bibnamefont
				{Villeneuve}}, \bibinfo {author} {\bibfnamefont {Andrei}\ \bibnamefont
				{Naumov}}, \bibinfo {author} {\bibfnamefont {Thomas}\ \bibnamefont {Brabec}},
			\ and\ \bibinfo {author} {\bibfnamefont {Giulio}\ \bibnamefont {Vampa}},\
		}\bibfield  {title} {\enquote {\bibinfo {title} {Photon bunching in
					high-harmonic emission controlled by quantum light},}\ }\href {\doibase
			10.1038/s41566-025-01673-6} {\bibfield  {journal} {\bibinfo  {journal}
				{Nature Photonics}\ }\textbf {\bibinfo {volume} {19}},\ \bibinfo {pages}
			{767--771} (\bibinfo {year} {2025})}\BibitemShut {NoStop}%
		\bibitem [{\citenamefont {Gorlach}\ \emph {et~al.}(2023)\citenamefont
			{Gorlach}, \citenamefont {Even~Tzur}, \citenamefont {Birk}, \citenamefont
			{Kr{\"u}ger}, \citenamefont {Rivera}, \citenamefont {Cohen},\ and\
			\citenamefont {Kaminer}}]{Gorlach2023HHGDrivenQuantumLight}%
		\BibitemOpen
		\bibfield  {author} {\bibinfo {author} {\bibfnamefont {Alexey}\ \bibnamefont
				{Gorlach}}, \bibinfo {author} {\bibfnamefont {Matan}\ \bibnamefont
				{Even~Tzur}}, \bibinfo {author} {\bibfnamefont {Michael}\ \bibnamefont
				{Birk}}, \bibinfo {author} {\bibfnamefont {Michael}\ \bibnamefont
				{Kr{\"u}ger}}, \bibinfo {author} {\bibfnamefont {Nicholas}\ \bibnamefont
				{Rivera}}, \bibinfo {author} {\bibfnamefont {Oren}\ \bibnamefont {Cohen}}, \
			and\ \bibinfo {author} {\bibfnamefont {Ido}\ \bibnamefont {Kaminer}},\
		}\bibfield  {title} {\enquote {\bibinfo {title} {High-harmonic generation
					driven by quantum light},}\ }\href {\doibase 10.1038/s41567-023-02127-y}
		{\bibfield  {journal} {\bibinfo  {journal} {Nature Physics}\ }\textbf
			{\bibinfo {volume} {19}},\ \bibinfo {pages} {1689--1696} (\bibinfo {year}
			{2023})}\BibitemShut {NoStop}%
		\bibitem [{\citenamefont {Rasputnyi}\ \emph {et~al.}(2024)\citenamefont
			{Rasputnyi}, \citenamefont {Chen}, \citenamefont {Birk}, \citenamefont
			{Cohen}, \citenamefont {Kaminer}, \citenamefont {Kr{\"u}ger}, \citenamefont
			{Seletskiy}, \citenamefont {Chekhova},\ and\ \citenamefont
			{Tani}}]{Rasputnyi2024HighHarmonicBSV}%
		\BibitemOpen
		\bibfield  {author} {\bibinfo {author} {\bibfnamefont {Andrei}\ \bibnamefont
				{Rasputnyi}}, \bibinfo {author} {\bibfnamefont {Zhaopin}\ \bibnamefont
				{Chen}}, \bibinfo {author} {\bibfnamefont {Michael}\ \bibnamefont {Birk}},
			\bibinfo {author} {\bibfnamefont {Oren}\ \bibnamefont {Cohen}}, \bibinfo
			{author} {\bibfnamefont {Ido}\ \bibnamefont {Kaminer}}, \bibinfo {author}
			{\bibfnamefont {Michael}\ \bibnamefont {Kr{\"u}ger}}, \bibinfo {author}
			{\bibfnamefont {Denis}\ \bibnamefont {Seletskiy}}, \bibinfo {author}
			{\bibfnamefont {Maria}\ \bibnamefont {Chekhova}}, \ and\ \bibinfo {author}
			{\bibfnamefont {Francesco}\ \bibnamefont {Tani}},\ }\bibfield  {title}
		{\enquote {\bibinfo {title} {High-harmonic generation by a bright squeezed
					vacuum},}\ }\href {\doibase 10.1038/s41567-024-02659-x} {\bibfield  {journal}
			{\bibinfo  {journal} {Nature Physics}\ }\textbf {\bibinfo {volume} {20}},\
			\bibinfo {pages} {1960--1965} (\bibinfo {year} {2024})}\BibitemShut {NoStop}%
		\bibitem [{\citenamefont {Heimerl}\ \emph {et~al.}(2024)\citenamefont
			{Heimerl}, \citenamefont {Mikhaylov}, \citenamefont {Meier}, \citenamefont
			{H{\"o}llerer}, \citenamefont {Kaminer}, \citenamefont {Chekhova},\ and\
			\citenamefont {Hommelhoff}}]{Heimerl2024MultiphotonElectronEmission}%
		\BibitemOpen
		\bibfield  {author} {\bibinfo {author} {\bibfnamefont {Jonas}\ \bibnamefont
				{Heimerl}}, \bibinfo {author} {\bibfnamefont {Alexander}\ \bibnamefont
				{Mikhaylov}}, \bibinfo {author} {\bibfnamefont {Stefan}\ \bibnamefont
				{Meier}}, \bibinfo {author} {\bibfnamefont {Henrick}\ \bibnamefont
				{H{\"o}llerer}}, \bibinfo {author} {\bibfnamefont {Ido}\ \bibnamefont
				{Kaminer}}, \bibinfo {author} {\bibfnamefont {Maria}\ \bibnamefont
				{Chekhova}}, \ and\ \bibinfo {author} {\bibfnamefont {Peter}\ \bibnamefont
				{Hommelhoff}},\ }\bibfield  {title} {\enquote {\bibinfo {title} {Multiphoton
					electron emission with non-classical light},}\ }\href {\doibase
			10.1038/s41567-024-02472-6} {\bibfield  {journal} {\bibinfo  {journal}
				{Nature Physics}\ }\textbf {\bibinfo {volume} {20}},\ \bibinfo {pages}
			{945--950} (\bibinfo {year} {2024})}\BibitemShut {NoStop}%
		\bibitem [{\citenamefont {Heimerl}\ \emph {et~al.}(2025)\citenamefont
			{Heimerl}, \citenamefont {Rasputnyi}, \citenamefont {P{\"o}lloth},
			\citenamefont {Meier}, \citenamefont {Chekhova},\ and\ \citenamefont
			{Hommelhoff}}]{Heimerl2025QuantumLightDrivesElectrons}%
		\BibitemOpen
		\bibfield  {author} {\bibinfo {author} {\bibfnamefont {Jonas}\ \bibnamefont
				{Heimerl}}, \bibinfo {author} {\bibfnamefont {Andrei}\ \bibnamefont
				{Rasputnyi}}, \bibinfo {author} {\bibfnamefont {Jonathan}\ \bibnamefont
				{P{\"o}lloth}}, \bibinfo {author} {\bibfnamefont {Stefan}\ \bibnamefont
				{Meier}}, \bibinfo {author} {\bibfnamefont {Maria}\ \bibnamefont {Chekhova}},
			\ and\ \bibinfo {author} {\bibfnamefont {Peter}\ \bibnamefont {Hommelhoff}},\
		}\bibfield  {title} {\enquote {\bibinfo {title} {Quantum light drives
					electrons strongly at metal needle tips},}\ }\href {\doibase
			10.1038/s41567-025-03087-1} {\bibfield  {journal} {\bibinfo  {journal}
				{Nature Physics}\ }\textbf {\bibinfo {volume} {21}},\ \bibinfo {pages}
			{1899--1904} (\bibinfo {year} {2025})}\BibitemShut {NoStop}%
		\bibitem [{\citenamefont {Even~Tzur}\ and\ \citenamefont
			{Cohen}(2024)}]{EvenTzur2024MotionChargedParticles}%
		\BibitemOpen
		\bibfield  {author} {\bibinfo {author} {\bibfnamefont {Matan}\ \bibnamefont
				{Even~Tzur}}\ and\ \bibinfo {author} {\bibfnamefont {Oren}\ \bibnamefont
				{Cohen}},\ }\bibfield  {title} {\enquote {\bibinfo {title} {Motion of charged
					particles in bright squeezed vacuum},}\ }\href {\doibase
			10.1038/s41377-024-01381-w} {\bibfield  {journal} {\bibinfo  {journal}
				{Light: Science \& Applications}\ }\textbf {\bibinfo {volume} {13}},\
			\bibinfo {pages} {41} (\bibinfo {year} {2024})}\BibitemShut {NoStop}%
		\bibitem [{\citenamefont {Fang}\ \emph {et~al.}(2023)\citenamefont {Fang},
			\citenamefont {Sun}, \citenamefont {He},\ and\ \citenamefont
			{Liu}}]{Fang2023StrongFieldIonization}%
		\BibitemOpen
		\bibfield  {author} {\bibinfo {author} {\bibfnamefont {Yiqi}\ \bibnamefont
				{Fang}}, \bibinfo {author} {\bibfnamefont {Fengxiao}\ \bibnamefont {Sun}},
			\bibinfo {author} {\bibfnamefont {Qiongyi}\ \bibnamefont {He}}, \ and\
			\bibinfo {author} {\bibfnamefont {Yunquan}\ \bibnamefont {Liu}},\ }\bibfield
		{title} {\enquote {\bibinfo {title} {Strong-field ionization of hydrogen
					atoms with quantum light},}\ }\href {\doibase 10.1103/PhysRevLett.130.253201}
		{\bibfield  {journal} {\bibinfo  {journal} {Physical Review Letters}\
			}\textbf {\bibinfo {volume} {130}},\ \bibinfo {pages} {253201} (\bibinfo
			{year} {2023})}\BibitemShut {NoStop}%
		\bibitem [{\citenamefont {Liu}\ \emph {et~al.}(2025)\citenamefont {Liu},
			\citenamefont {Zhang}, \citenamefont {Wang},\ and\ \citenamefont
			{Yuan}}]{Liu2025AtomicDoubleIonizationQuantumLight}%
		\BibitemOpen
		\bibfield  {author} {\bibinfo {author} {\bibfnamefont {Haoyu}\ \bibnamefont
				{Liu}}, \bibinfo {author} {\bibfnamefont {Hanxu}\ \bibnamefont {Zhang}},
			\bibinfo {author} {\bibfnamefont {Xu}~\bibnamefont {Wang}}, \ and\ \bibinfo
			{author} {\bibfnamefont {Jianmin}\ \bibnamefont {Yuan}},\ }\bibfield  {title}
		{\enquote {\bibinfo {title} {Atomic double ionization with quantum light},}\
		}\href {\doibase 10.1103/PhysRevLett.134.123202} {\bibfield  {journal}
			{\bibinfo  {journal} {Physical Review Letters}\ }\textbf {\bibinfo {volume}
				{134}},\ \bibinfo {pages} {123202} (\bibinfo {year} {2025})}\BibitemShut
		{NoStop}%
		\bibitem [{\citenamefont {Sh.~Iskhakov}\ \emph {et~al.}(2012)\citenamefont
			{Sh.~Iskhakov}, \citenamefont {P{\'e}rez}, \citenamefont {Yu.~Spasibko},
			\citenamefont {Chekhova},\ and\ \citenamefont {Leuchs}}]{Ishkhakov2012}%
		\BibitemOpen
		\bibfield  {author} {\bibinfo {author} {\bibfnamefont {T}~\bibnamefont
				{Sh.~Iskhakov}}, \bibinfo {author} {\bibfnamefont {AM}~\bibnamefont
				{P{\'e}rez}}, \bibinfo {author} {\bibfnamefont {K}~\bibnamefont
				{Yu.~Spasibko}}, \bibinfo {author} {\bibfnamefont {MV}~\bibnamefont
				{Chekhova}}, \ and\ \bibinfo {author} {\bibfnamefont {G}~\bibnamefont
				{Leuchs}},\ }\bibfield  {title} {\enquote {\bibinfo {title} {Superbunched
					bright squeezed vacuum state},}\ }\href {\doibase 10.1364/OL.37.001919}
		{\bibfield  {journal} {\bibinfo  {journal} {Optics Letters}\ }\textbf
			{\bibinfo {volume} {37}},\ \bibinfo {pages} {1919--1921} (\bibinfo {year}
			{2012})}\BibitemShut {NoStop}%
		\bibitem [{\citenamefont {Finger}\ \emph {et~al.}(2015)\citenamefont {Finger},
			\citenamefont {Iskhakov}, \citenamefont {Joly}, \citenamefont {Chekhova},\
			and\ \citenamefont {Russell}}]{Finger2015}%
		\BibitemOpen
		\bibfield  {author} {\bibinfo {author} {\bibfnamefont {Martin~A}\
				\bibnamefont {Finger}}, \bibinfo {author} {\bibfnamefont {Timur~Sh}\
				\bibnamefont {Iskhakov}}, \bibinfo {author} {\bibfnamefont {Nicolas~Y}\
				\bibnamefont {Joly}}, \bibinfo {author} {\bibfnamefont {Maria~V}\
				\bibnamefont {Chekhova}}, \ and\ \bibinfo {author} {\bibfnamefont {Philip
					St~J}\ \bibnamefont {Russell}},\ }\bibfield  {title} {\enquote {\bibinfo
				{title} {Raman-free, noble-gas-filled photonic-crystal fiber source for
					ultrafast, very bright twin-beam squeezed vacuum},}\ }\href {\doibase
			10.1103/PhysRevLett.115.143602} {\bibfield  {journal} {\bibinfo  {journal}
				{Physical Review Letters}\ }\textbf {\bibinfo {volume} {115}},\ \bibinfo
			{pages} {143602} (\bibinfo {year} {2015})}\BibitemShut {NoStop}%
		\bibitem [{\citenamefont {Schleich}(2001)}]{Schleich2001}%
		\BibitemOpen
		\bibfield  {author} {\bibinfo {author} {\bibfnamefont {Wolfgang~P.}\
				\bibnamefont {Schleich}},\ }\href@noop {} {\emph {\bibinfo {title} {Quantum
					Optics in Phase Space}}}\ (\bibinfo  {publisher} {Wiley-VCH},\ \bibinfo
		{address} {Berlin},\ \bibinfo {year} {2001})\BibitemShut {NoStop}%
		\bibitem [{\citenamefont {Kern}\ \emph {et~al.}(2026)\citenamefont {Kern},
			\citenamefont {Nisim}, \citenamefont {Birk}, \citenamefont {Rasputnyi},
			\citenamefont {Behar}, \citenamefont {Chen}, \citenamefont {Kaminer},
			\citenamefont {Sidorenko}, \citenamefont {Cohen},\ and\ \citenamefont
			{Kr{\"u}ger}}]{Kern2026Optica}%
		\BibitemOpen
		\bibfield  {author} {\bibinfo {author} {\bibfnamefont {Yuval}\ \bibnamefont
				{Kern}}, \bibinfo {author} {\bibfnamefont {Ido}\ \bibnamefont {Nisim}},
			\bibinfo {author} {\bibfnamefont {Michael}\ \bibnamefont {Birk}}, \bibinfo
			{author} {\bibfnamefont {Andrei}\ \bibnamefont {Rasputnyi}}, \bibinfo
			{author} {\bibfnamefont {Doron}\ \bibnamefont {Behar}}, \bibinfo {author}
			{\bibfnamefont {Zhaopin}\ \bibnamefont {Chen}}, \bibinfo {author}
			{\bibfnamefont {Ido}\ \bibnamefont {Kaminer}}, \bibinfo {author}
			{\bibfnamefont {Pavel}\ \bibnamefont {Sidorenko}}, \bibinfo {author}
			{\bibfnamefont {Oren}\ \bibnamefont {Cohen}}, \ and\ \bibinfo {author}
			{\bibfnamefont {Michael}\ \bibnamefont {Kr{\"u}ger}},\ }\bibfield  {title}
		{\enquote {\bibinfo {title} {Single-shot pulse retrieval of femtosecond
					bright squeezed vacuum},}\ }\href {\doibase 10.1364/OPTICA.580767} {\bibfield
			{journal} {\bibinfo  {journal} {Optica}\ }\textbf {\bibinfo {volume} {13}},\
			\bibinfo {pages} {395--399} (\bibinfo {year} {2026})}\BibitemShut {NoStop}%
		\bibitem [{\citenamefont {Rabi}(1937)}]{Rabi1937}%
		\BibitemOpen
		\bibfield  {author} {\bibinfo {author} {\bibfnamefont {I.~I.}\ \bibnamefont
				{Rabi}},\ }\bibfield  {title} {\enquote {\bibinfo {title} {Space quantization
					in a gyrating magnetic field},}\ }\href {\doibase 10.1103/PhysRev.51.652}
		{\bibfield  {journal} {\bibinfo  {journal} {Phys. Rev.}\ }\textbf {\bibinfo
				{volume} {51}},\ \bibinfo {pages} {652--654} (\bibinfo {year}
			{1937})}\BibitemShut {NoStop}%
		\bibitem [{\citenamefont {Knight}\ and\ \citenamefont
			{Milonni}(1980)}]{Knight1980}%
		\BibitemOpen
		\bibfield  {author} {\bibinfo {author} {\bibfnamefont {P.L.}\ \bibnamefont
				{Knight}}\ and\ \bibinfo {author} {\bibfnamefont {P.W.}\ \bibnamefont
				{Milonni}},\ }\bibfield  {title} {\enquote {\bibinfo {title} {The rabi
					frequency in optical spectra},}\ }\href {\doibase
			https://doi.org/10.1016/0370-1573(80)90119-2} {\bibfield  {journal} {\bibinfo
				{journal} {Physics Reports}\ }\textbf {\bibinfo {volume} {66}},\ \bibinfo
			{pages} {21--107} (\bibinfo {year} {1980})}\BibitemShut {NoStop}%
		\bibitem [{\citenamefont {Walker}\ \emph {et~al.}(1995)\citenamefont {Walker},
			\citenamefont {Kalu\ifmmode~\check{z}\else \v{z}\fi{}a}, \citenamefont
			{Sheehy}, \citenamefont {Agostini},\ and\ \citenamefont
			{DiMauro}}]{Walker1995}%
		\BibitemOpen
		\bibfield  {author} {\bibinfo {author} {\bibfnamefont {Barry}\ \bibnamefont
				{Walker}}, \bibinfo {author} {\bibfnamefont {M.}~\bibnamefont
				{Kalu\ifmmode~\check{z}\else \v{z}\fi{}a}}, \bibinfo {author} {\bibfnamefont
				{B.}~\bibnamefont {Sheehy}}, \bibinfo {author} {\bibfnamefont
				{P.}~\bibnamefont {Agostini}}, \ and\ \bibinfo {author} {\bibfnamefont
				{L.~F.}\ \bibnamefont {DiMauro}},\ }\bibfield  {title} {\enquote {\bibinfo
				{title} {Observation of continuum-continuum autler-townes splitting},}\
		}\href {\doibase 10.1103/PhysRevLett.75.633} {\bibfield  {journal} {\bibinfo
				{journal} {Phys. Rev. Lett.}\ }\textbf {\bibinfo {volume} {75}},\ \bibinfo
			{pages} {633--636} (\bibinfo {year} {1995})}\BibitemShut {NoStop}%
		\bibitem [{\citenamefont {Sun}\ and\ \citenamefont {Lou}(2003)}]{Sun2003}%
		\BibitemOpen
		\bibfield  {author} {\bibinfo {author} {\bibfnamefont {Zhigang}\ \bibnamefont
				{Sun}}\ and\ \bibinfo {author} {\bibfnamefont {Nanquan}\ \bibnamefont
				{Lou}},\ }\bibfield  {title} {\enquote {\bibinfo {title} {Autler-townes
					splitting in the multiphoton resonance ionization spectrum of molecules
					produced by ultrashort laser pulses},}\ }\href {\doibase
			10.1103/PhysRevLett.91.023002} {\bibfield  {journal} {\bibinfo  {journal}
				{Phys. Rev. Lett.}\ }\textbf {\bibinfo {volume} {91}},\ \bibinfo {pages}
			{023002} (\bibinfo {year} {2003})}\BibitemShut {NoStop}%
		\bibitem [{\citenamefont {Wollenhaupt}\ \emph {et~al.}(2003)\citenamefont
			{Wollenhaupt}, \citenamefont {Assion}, \citenamefont {Bazhan}, \citenamefont
			{Horn}, \citenamefont {Liese}, \citenamefont {Sarpe-Tudoran}, \citenamefont
			{Winter},\ and\ \citenamefont {Baumert}}]{Wollenhaupt2003}%
		\BibitemOpen
		\bibfield  {author} {\bibinfo {author} {\bibfnamefont {M.}~\bibnamefont
				{Wollenhaupt}}, \bibinfo {author} {\bibfnamefont {A.}~\bibnamefont {Assion}},
			\bibinfo {author} {\bibfnamefont {O.}~\bibnamefont {Bazhan}}, \bibinfo
			{author} {\bibfnamefont {Ch.}\ \bibnamefont {Horn}}, \bibinfo {author}
			{\bibfnamefont {D.}~\bibnamefont {Liese}}, \bibinfo {author} {\bibfnamefont
				{Ch.}\ \bibnamefont {Sarpe-Tudoran}}, \bibinfo {author} {\bibfnamefont
				{M.}~\bibnamefont {Winter}}, \ and\ \bibinfo {author} {\bibfnamefont
				{T.}~\bibnamefont {Baumert}},\ }\bibfield  {title} {\enquote {\bibinfo
				{title} {Control of interferences in an autler-townes doublet: Symmetry of
					control parameters},}\ }\href {\doibase 10.1103/PhysRevA.68.015401}
		{\bibfield  {journal} {\bibinfo  {journal} {Physical Review A}\ }\textbf
			{\bibinfo {volume} {68}},\ \bibinfo {pages} {015401} (\bibinfo {year}
			{2003})}\BibitemShut {NoStop}%
		\bibitem [{\citenamefont {Kaiser}\ \emph {et~al.}(2013)\citenamefont {Kaiser},
			\citenamefont {Brand}, \citenamefont {Glässl}, \citenamefont {Vagov},
			\citenamefont {Axt},\ and\ \citenamefont {Pietsch}}]{Kaiser2013}%
		\BibitemOpen
		\bibfield  {author} {\bibinfo {author} {\bibfnamefont {B}~\bibnamefont
				{Kaiser}}, \bibinfo {author} {\bibfnamefont {A}~\bibnamefont {Brand}},
			\bibinfo {author} {\bibfnamefont {M}~\bibnamefont {Glässl}}, \bibinfo
			{author} {\bibfnamefont {A}~\bibnamefont {Vagov}}, \bibinfo {author}
			{\bibfnamefont {V~M}\ \bibnamefont {Axt}}, \ and\ \bibinfo {author}
			{\bibfnamefont {U}~\bibnamefont {Pietsch}},\ }\bibfield  {title} {\enquote
			{\bibinfo {title} {Photoionization of resonantly driven atomic states by an
					extreme ultraviolet-free-electron laser: intensity dependence and
					renormalization of rabi frequencies},}\ }\href {\doibase
			10.1088/1367-2630/15/9/093016} {\bibfield  {journal} {\bibinfo  {journal}
				{New Journal of Physics}\ }\textbf {\bibinfo {volume} {15}},\ \bibinfo
			{pages} {093016} (\bibinfo {year} {2013})}\BibitemShut {NoStop}%
		\bibitem [{\citenamefont {Ciappina}\ \emph {et~al.}(2015)\citenamefont
			{Ciappina}, \citenamefont {P{\'e}rez-Hern{\'a}ndez}, \citenamefont
			{Landsman}, \citenamefont {Zimmermann}, \citenamefont {Lewenstein},
			\citenamefont {Roso},\ and\ \citenamefont {Krausz}}]{Ciappina2015}%
		\BibitemOpen
		\bibfield  {author} {\bibinfo {author} {\bibfnamefont {M.~F.}\ \bibnamefont
				{Ciappina}}, \bibinfo {author} {\bibfnamefont {J.~A.}\ \bibnamefont
				{P{\'e}rez-Hern{\'a}ndez}}, \bibinfo {author} {\bibfnamefont {A.~S.}\
				\bibnamefont {Landsman}}, \bibinfo {author} {\bibfnamefont {T.}~\bibnamefont
				{Zimmermann}}, \bibinfo {author} {\bibfnamefont {M.}~\bibnamefont
				{Lewenstein}}, \bibinfo {author} {\bibfnamefont {L.}~\bibnamefont {Roso}}, \
			and\ \bibinfo {author} {\bibfnamefont {F.}~\bibnamefont {Krausz}},\
		}\bibfield  {title} {\enquote {\bibinfo {title} {Carrier-wave rabi-flopping
					signatures in high-order harmonic generation for alkali atoms},}\ }\href
		{\doibase 10.1103/PhysRevLett.114.143902} {\bibfield  {journal} {\bibinfo
				{journal} {Physical Review Letters}\ }\textbf {\bibinfo {volume} {114}},\
			\bibinfo {pages} {143902} (\bibinfo {year} {2015})}\BibitemShut {NoStop}%
		\bibitem [{\citenamefont {Fushitani}\ \emph {et~al.}(2016)\citenamefont
			{Fushitani}, \citenamefont {Liu}, \citenamefont {Matsuda}, \citenamefont
			{Endo}, \citenamefont {Toida}, \citenamefont {Nagasono}, \citenamefont
			{Togashi}, \citenamefont {Yabashi}, \citenamefont {Ishikawa}, \citenamefont
			{Hikosaka} \emph {et~al.}}]{Fushitani2015}%
		\BibitemOpen
		\bibfield  {author} {\bibinfo {author} {\bibfnamefont {M}~\bibnamefont
				{Fushitani}}, \bibinfo {author} {\bibfnamefont {C-N}\ \bibnamefont {Liu}},
			\bibinfo {author} {\bibfnamefont {A}~\bibnamefont {Matsuda}}, \bibinfo
			{author} {\bibfnamefont {T}~\bibnamefont {Endo}}, \bibinfo {author}
			{\bibfnamefont {Y}~\bibnamefont {Toida}}, \bibinfo {author} {\bibfnamefont
				{M}~\bibnamefont {Nagasono}}, \bibinfo {author} {\bibfnamefont
				{T}~\bibnamefont {Togashi}}, \bibinfo {author} {\bibfnamefont
				{M}~\bibnamefont {Yabashi}}, \bibinfo {author} {\bibfnamefont
				{T}~\bibnamefont {Ishikawa}}, \bibinfo {author} {\bibfnamefont
				{Y}~\bibnamefont {Hikosaka}},  \emph {et~al.},\ }\bibfield  {title} {\enquote
			{\bibinfo {title} {Femtosecond two-photon rabi oscillations in excited he
					driven by ultrashort intense laser fields},}\ }\href {\doibase
			10.1038/nphoton.2015.228} {\bibfield  {journal} {\bibinfo  {journal} {Nature
					Photonics}\ }\textbf {\bibinfo {volume} {10}},\ \bibinfo {pages} {102--105}
			(\bibinfo {year} {2016})}\BibitemShut {NoStop}%
		\bibitem [{\citenamefont {Tumakov}\ \emph {et~al.}(2019)\citenamefont
			{Tumakov}, \citenamefont {Telnov}, \citenamefont {Plunien},\ and\
			\citenamefont {Shabaev}}]{Tumakov2019}%
		\BibitemOpen
		\bibfield  {author} {\bibinfo {author} {\bibfnamefont {D.~A.}\ \bibnamefont
				{Tumakov}}, \bibinfo {author} {\bibfnamefont {Dmitry~A.}\ \bibnamefont
				{Telnov}}, \bibinfo {author} {\bibfnamefont {G.}~\bibnamefont {Plunien}}, \
			and\ \bibinfo {author} {\bibfnamefont {V.~M.}\ \bibnamefont {Shabaev}},\
		}\bibfield  {title} {\enquote {\bibinfo {title} {Photoelectron spectra after
					multiphoton ionization of li atoms in the one-photon rabi-flopping regime},}\
		}\href {\doibase 10.1103/PhysRevA.100.023407} {\bibfield  {journal} {\bibinfo
				{journal} {Phys. Rev. A}\ }\textbf {\bibinfo {volume} {100}},\ \bibinfo
			{pages} {023407} (\bibinfo {year} {2019})}\BibitemShut {NoStop}%
		\bibitem [{\citenamefont {Li}\ \emph {et~al.}(2021)\citenamefont {Li},
			\citenamefont {Lei}, \citenamefont {Li}, \citenamefont {Yang}, \citenamefont
			{Du}, \citenamefont {Jiang}, \citenamefont {Li}, \citenamefont {Liu},
			\citenamefont {He}, \citenamefont {Ma} \emph {et~al.}}]{LiWankai2021}%
		\BibitemOpen
		\bibfield  {author} {\bibinfo {author} {\bibfnamefont {Wankai}\ \bibnamefont
				{Li}}, \bibinfo {author} {\bibfnamefont {Yue}\ \bibnamefont {Lei}}, \bibinfo
			{author} {\bibfnamefont {Xing}\ \bibnamefont {Li}}, \bibinfo {author}
			{\bibfnamefont {Tao}\ \bibnamefont {Yang}}, \bibinfo {author} {\bibfnamefont
				{Mei}\ \bibnamefont {Du}}, \bibinfo {author} {\bibfnamefont {Ying}\
				\bibnamefont {Jiang}}, \bibinfo {author} {\bibfnamefont {Jialong}\
				\bibnamefont {Li}}, \bibinfo {author} {\bibfnamefont {Aihua}\ \bibnamefont
				{Liu}}, \bibinfo {author} {\bibfnamefont {Lanhai}\ \bibnamefont {He}},
			\bibinfo {author} {\bibfnamefont {Pan}\ \bibnamefont {Ma}},  \emph {et~al.},\
		}\bibfield  {title} {\enquote {\bibinfo {title} {New source for tuning the
					effective rabi frequency discovered in multiphoton ionization},}\ }\href
		{https://doi.org/10.48550/arXiv.2112.13096} {\bibfield  {journal} {\bibinfo
				{journal} {arXiv preprint arXiv:2112.13096}\ } (\bibinfo {year}
			{2021})}\BibitemShut {NoStop}%
		\bibitem [{\citenamefont {T{\'{o}}th}\ and\ \citenamefont
			{Csehi}(2021)}]{Csehi2021}%
		\BibitemOpen
		\bibfield  {author} {\bibinfo {author} {\bibfnamefont {Attila}\ \bibnamefont
				{T{\'{o}}th}}\ and\ \bibinfo {author} {\bibfnamefont {Andr{\'{a}}s}\
				\bibnamefont {Csehi}},\ }\bibfield  {title} {\enquote {\bibinfo {title}
				{Probing strong-field two-photon transitions through dynamic interference},}\
		}\href {\doibase 10.1088/1361-6455/abdb8e} {\bibfield  {journal} {\bibinfo
				{journal} {Journal of Physics B: Atomic, Molecular and Optical Physics}\
			}\textbf {\bibinfo {volume} {54}},\ \bibinfo {pages} {035005} (\bibinfo
			{year} {2021})}\BibitemShut {NoStop}%
		\bibitem [{\citenamefont {Jiang}\ \emph {et~al.}(2021)\citenamefont {Jiang},
			\citenamefont {Liang}, \citenamefont {Wang}, \citenamefont {Peng},\ and\
			\citenamefont {Burgd\"orfer}}]{Jiang2021}%
		\BibitemOpen
		\bibfield  {author} {\bibinfo {author} {\bibfnamefont {Wei-Chao}\
				\bibnamefont {Jiang}}, \bibinfo {author} {\bibfnamefont {Hao}\ \bibnamefont
				{Liang}}, \bibinfo {author} {\bibfnamefont {Shun}\ \bibnamefont {Wang}},
			\bibinfo {author} {\bibfnamefont {Liang-You}\ \bibnamefont {Peng}}, \ and\
			\bibinfo {author} {\bibfnamefont {Joachim}\ \bibnamefont {Burgd\"orfer}},\
		}\bibfield  {title} {\enquote {\bibinfo {title} {Enhancing autler-townes
					splittings by ultrafast xuv pulses},}\ }\href {\doibase
			10.1103/PhysRevResearch.3.L032052} {\bibfield  {journal} {\bibinfo  {journal}
				{Phys. Rev. Research}\ }\textbf {\bibinfo {volume} {3}},\ \bibinfo {pages}
			{L032052} (\bibinfo {year} {2021})}\BibitemShut {NoStop}%
		\bibitem [{\citenamefont {Zhang}\ \emph {et~al.}(2022)\citenamefont {Zhang},
			\citenamefont {Zhou}, \citenamefont {Liao}, \citenamefont {Chen},
			\citenamefont {Liang}, \citenamefont {Ke}, \citenamefont {Li}, \citenamefont
			{Csehi},\ and\ \citenamefont {Lu}}]{Zhang2022}%
		\BibitemOpen
		\bibfield  {author} {\bibinfo {author} {\bibfnamefont {Xu}~\bibnamefont
				{Zhang}}, \bibinfo {author} {\bibfnamefont {Yueming}\ \bibnamefont {Zhou}},
			\bibinfo {author} {\bibfnamefont {Yijie}\ \bibnamefont {Liao}}, \bibinfo
			{author} {\bibfnamefont {Yongkun}\ \bibnamefont {Chen}}, \bibinfo {author}
			{\bibfnamefont {Jintai}\ \bibnamefont {Liang}}, \bibinfo {author}
			{\bibfnamefont {Qinghua}\ \bibnamefont {Ke}}, \bibinfo {author}
			{\bibfnamefont {Min}\ \bibnamefont {Li}}, \bibinfo {author} {\bibfnamefont
				{Andr\'as}\ \bibnamefont {Csehi}}, \ and\ \bibinfo {author} {\bibfnamefont
				{Peixiang}\ \bibnamefont {Lu}},\ }\bibfield  {title} {\enquote {\bibinfo
				{title} {Effect of nonresonant states in near-resonant two-photon ionization
					of hydrogen},}\ }\href {\doibase 10.1103/PhysRevA.106.063114} {\bibfield
			{journal} {\bibinfo  {journal} {Phys. Rev. A}\ }\textbf {\bibinfo {volume}
				{106}},\ \bibinfo {pages} {063114} (\bibinfo {year} {2022})}\BibitemShut
		{NoStop}%
		\bibitem [{\citenamefont {Nandi}\ \emph {et~al.}(2022)\citenamefont {Nandi},
			\citenamefont {Olofsson}, \citenamefont {Bertolino}, \citenamefont
			{Carlstr{\"o}m}, \citenamefont {Zapata}, \citenamefont {Busto}, \citenamefont
			{Callegari}, \citenamefont {Di~Fraia}, \citenamefont {Eng-Johnsson},
			\citenamefont {Feifel} \emph {et~al.}}]{Nandi2022}%
		\BibitemOpen
		\bibfield  {author} {\bibinfo {author} {\bibfnamefont {Saikat}\ \bibnamefont
				{Nandi}}, \bibinfo {author} {\bibfnamefont {Edvin}\ \bibnamefont {Olofsson}},
			\bibinfo {author} {\bibfnamefont {Mattias}\ \bibnamefont {Bertolino}},
			\bibinfo {author} {\bibfnamefont {Stefanos}\ \bibnamefont {Carlstr{\"o}m}},
			\bibinfo {author} {\bibfnamefont {Felipe}\ \bibnamefont {Zapata}}, \bibinfo
			{author} {\bibfnamefont {David}\ \bibnamefont {Busto}}, \bibinfo {author}
			{\bibfnamefont {Carlo}\ \bibnamefont {Callegari}}, \bibinfo {author}
			{\bibfnamefont {Michele}\ \bibnamefont {Di~Fraia}}, \bibinfo {author}
			{\bibfnamefont {Per}\ \bibnamefont {Eng-Johnsson}}, \bibinfo {author}
			{\bibfnamefont {Raimund}\ \bibnamefont {Feifel}},  \emph {et~al.},\
		}\bibfield  {title} {\enquote {\bibinfo {title} {Observation of rabi dynamics
					with a short-wavelength free-electron laser},}\ }\href {\doibase
			10.1038/s41586-022-04948-y} {\bibfield  {journal} {\bibinfo  {journal}
				{Nature}\ }\textbf {\bibinfo {volume} {608}},\ \bibinfo {pages} {488--493}
			(\bibinfo {year} {2022})}\BibitemShut {NoStop}%
		\bibitem [{\citenamefont {Liao}\ \emph {et~al.}(2022)\citenamefont {Liao},
			\citenamefont {Zhou}, \citenamefont {Pi}, \citenamefont {Liang},
			\citenamefont {Ke}, \citenamefont {Zhao}, \citenamefont {Li},\ and\
			\citenamefont {Lu}}]{Liao2022}%
		\BibitemOpen
		\bibfield  {author} {\bibinfo {author} {\bibfnamefont {Yijie}\ \bibnamefont
				{Liao}}, \bibinfo {author} {\bibfnamefont {Yueming}\ \bibnamefont {Zhou}},
			\bibinfo {author} {\bibfnamefont {Liang-Wen}\ \bibnamefont {Pi}}, \bibinfo
			{author} {\bibfnamefont {Jintai}\ \bibnamefont {Liang}}, \bibinfo {author}
			{\bibfnamefont {Qinghua}\ \bibnamefont {Ke}}, \bibinfo {author}
			{\bibfnamefont {Yong}\ \bibnamefont {Zhao}}, \bibinfo {author} {\bibfnamefont
				{Min}\ \bibnamefont {Li}}, \ and\ \bibinfo {author} {\bibfnamefont
				{Peixiang}\ \bibnamefont {Lu}},\ }\bibfield  {title} {\enquote {\bibinfo
				{title} {Reconstruction of attosecond beating by interference of two-photon
					transitions on the lithium atom with rabi oscillations},}\ }\href {\doibase
			10.1103/PhysRevA.105.063110} {\bibfield  {journal} {\bibinfo  {journal}
				{Phys. Rev. A}\ }\textbf {\bibinfo {volume} {105}},\ \bibinfo {pages}
			{063110} (\bibinfo {year} {2022})}\BibitemShut {NoStop}%
		\bibitem [{\citenamefont {Olofsson}\ and\ \citenamefont
			{Dahlstr{\"o}m}(2023)}]{olofsson2023}%
		\BibitemOpen
		\bibfield  {author} {\bibinfo {author} {\bibfnamefont {Edvin}\ \bibnamefont
				{Olofsson}}\ and\ \bibinfo {author} {\bibfnamefont {Jan~Marcus}\ \bibnamefont
				{Dahlstr{\"o}m}},\ }\bibfield  {title} {\enquote {\bibinfo {title}
				{Photoelectron signature of dressed-atom stabilization in an intense xuv
					field},}\ }\href {\doibase 10.1103/PhysRevResearch.5.043017} {\bibfield
			{journal} {\bibinfo  {journal} {Physical Review Research}\ }\textbf {\bibinfo
				{volume} {5}},\ \bibinfo {pages} {043017} (\bibinfo {year}
			{2023})}\BibitemShut {NoStop}%
		\bibitem [{\citenamefont {Bayer}\ \emph {et~al.}(2023)\citenamefont {Bayer},
			\citenamefont {Eickhoff}, \citenamefont {K{\"o}hnke},\ and\ \citenamefont
			{Wollenhaupt}}]{Wollenhaupt2023}%
		\BibitemOpen
		\bibfield  {author} {\bibinfo {author} {\bibfnamefont {T.}~\bibnamefont
				{Bayer}}, \bibinfo {author} {\bibfnamefont {K.}~\bibnamefont {Eickhoff}},
			\bibinfo {author} {\bibfnamefont {D.}~\bibnamefont {K{\"o}hnke}}, \ and\
			\bibinfo {author} {\bibfnamefont {M.}~\bibnamefont {Wollenhaupt}},\
		}\bibfield  {title} {\enquote {\bibinfo {title} {Phase control of the
					autler-townes doublet in multistate systems},}\ }\href {\doibase
			10.1103/PhysRevA.108.033111} {\bibfield  {journal} {\bibinfo  {journal}
				{Physical Review A}\ }\textbf {\bibinfo {volume} {108}},\ \bibinfo {pages}
			{033111} (\bibinfo {year} {2023})}\BibitemShut {NoStop}%
		\bibitem [{\citenamefont {Cui}\ \emph {et~al.}(2023)\citenamefont {Cui},
			\citenamefont {Cheng}, \citenamefont {Wang}, \citenamefont {Li},
			\citenamefont {Rohringer}, \citenamefont {Kimberg},\ and\ \citenamefont
			{Zhang}}]{Cui2023}%
		\BibitemOpen
		\bibfield  {author} {\bibinfo {author} {\bibfnamefont {Jun~Jie}\ \bibnamefont
				{Cui}}, \bibinfo {author} {\bibfnamefont {Yongjun}\ \bibnamefont {Cheng}},
			\bibinfo {author} {\bibfnamefont {Xin}\ \bibnamefont {Wang}}, \bibinfo
			{author} {\bibfnamefont {Zheng}\ \bibnamefont {Li}}, \bibinfo {author}
			{\bibfnamefont {Nina}\ \bibnamefont {Rohringer}}, \bibinfo {author}
			{\bibfnamefont {Victor}\ \bibnamefont {Kimberg}}, \ and\ \bibinfo {author}
			{\bibfnamefont {Song~Bin}\ \bibnamefont {Zhang}},\ }\bibfield  {title}
		{\enquote {\bibinfo {title} {Proposal for observing xuv-induced rabi
					oscillation using superfluorescent emission},}\ }\href {\doibase
			10.1103/PhysRevLett.131.043201} {\bibfield  {journal} {\bibinfo  {journal}
				{Physical Review Letters}\ }\textbf {\bibinfo {volume} {131}},\ \bibinfo
			{pages} {043201} (\bibinfo {year} {2023})}\BibitemShut {NoStop}%
		\bibitem [{\citenamefont {Pan}\ \emph {et~al.}(2023)\citenamefont {Pan},
			\citenamefont {Hu}, \citenamefont {Zhang}, \citenamefont {Zhang},
			\citenamefont {Zhou}, \citenamefont {Lu}, \citenamefont {Lu}, \citenamefont
			{Ni}, \citenamefont {Wu},\ and\ \citenamefont {He}}]{Pan2023}%
		\BibitemOpen
		\bibfield  {author} {\bibinfo {author} {\bibfnamefont {Shengzhe}\
				\bibnamefont {Pan}}, \bibinfo {author} {\bibfnamefont {Chenxi}\ \bibnamefont
				{Hu}}, \bibinfo {author} {\bibfnamefont {Wenbin}\ \bibnamefont {Zhang}},
			\bibinfo {author} {\bibfnamefont {Zhaohan}\ \bibnamefont {Zhang}}, \bibinfo
			{author} {\bibfnamefont {Lianrong}\ \bibnamefont {Zhou}}, \bibinfo {author}
			{\bibfnamefont {Chenxu}\ \bibnamefont {Lu}}, \bibinfo {author} {\bibfnamefont
				{Peifen}\ \bibnamefont {Lu}}, \bibinfo {author} {\bibfnamefont {Hongcheng}\
				\bibnamefont {Ni}}, \bibinfo {author} {\bibfnamefont {Jian}\ \bibnamefont
				{Wu}}, \ and\ \bibinfo {author} {\bibfnamefont {Feng}\ \bibnamefont {He}},\
		}\bibfield  {title} {\enquote {\bibinfo {title} {Rabi oscillations in a
					stretching molecule},}\ }\href {\doibase 10.1038/s41377-023-01075-9}
		{\bibfield  {journal} {\bibinfo  {journal} {Light: Science \& Applications}\
			}\textbf {\bibinfo {volume} {12}},\ \bibinfo {pages} {35} (\bibinfo {year}
			{2023})}\BibitemShut {NoStop}%
		\bibitem [{\citenamefont {Ishikawa}\ \emph {et~al.}(2023)\citenamefont
			{Ishikawa}, \citenamefont {Prince},\ and\ \citenamefont
			{Ueda}}]{Ishikawa2023}%
		\BibitemOpen
		\bibfield  {author} {\bibinfo {author} {\bibfnamefont {Kenichi~L}\
				\bibnamefont {Ishikawa}}, \bibinfo {author} {\bibfnamefont {Kevin~C}\
				\bibnamefont {Prince}}, \ and\ \bibinfo {author} {\bibfnamefont {Kiyoshi}\
				\bibnamefont {Ueda}},\ }\bibfield  {title} {\enquote {\bibinfo {title}
				{Control of ion-photoelectron entanglement and coherence via rabi
					oscillations},}\ }\href {\doibase 10.1021/acs.jpca.3c06781} {\bibfield
			{journal} {\bibinfo  {journal} {The Journal of Physical Chemistry A}\
			}\textbf {\bibinfo {volume} {127}},\ \bibinfo {pages} {10638--10646}
			(\bibinfo {year} {2023})}\BibitemShut {NoStop}%
		\bibitem [{\citenamefont {Nandi}\ \emph {et~al.}(2024)\citenamefont {Nandi},
			\citenamefont {Stenquist}, \citenamefont {Papoulia}, \citenamefont
			{Olofsson}, \citenamefont {Badano}, \citenamefont {Bertolino}, \citenamefont
			{Busto}, \citenamefont {Callegari}, \citenamefont {Carlström}, \citenamefont
			{Danailov}, \citenamefont {Demekhin}, \citenamefont {Fraia}, \citenamefont
			{Eng-Johnsson}, \citenamefont {Feifel}, \citenamefont {Gallician},
			\citenamefont {Giannessi}, \citenamefont {Gisselbrecht}, \citenamefont
			{Manfredda}, \citenamefont {Meyer}, \citenamefont {Miron}, \citenamefont
			{Peschel}, \citenamefont {Plekan}, \citenamefont {Prince}, \citenamefont
			{Squibb}, \citenamefont {Zangrando}, \citenamefont {Zapata}, \citenamefont
			{Zhong},\ and\ \citenamefont {Dahlström}}]{Nandi2024}%
		\BibitemOpen
		\bibfield  {author} {\bibinfo {author} {\bibfnamefont {Saikat}\ \bibnamefont
				{Nandi}}, \bibinfo {author} {\bibfnamefont {Axel}\ \bibnamefont {Stenquist}},
			\bibinfo {author} {\bibfnamefont {Asimina}\ \bibnamefont {Papoulia}},
			\bibinfo {author} {\bibfnamefont {Edvin}\ \bibnamefont {Olofsson}}, \bibinfo
			{author} {\bibfnamefont {Laura}\ \bibnamefont {Badano}}, \bibinfo {author}
			{\bibfnamefont {Mattias}\ \bibnamefont {Bertolino}}, \bibinfo {author}
			{\bibfnamefont {David}\ \bibnamefont {Busto}}, \bibinfo {author}
			{\bibfnamefont {Carlo}\ \bibnamefont {Callegari}}, \bibinfo {author}
			{\bibfnamefont {Stefanos}\ \bibnamefont {Carlström}}, \bibinfo {author}
			{\bibfnamefont {Miltcho~B.}\ \bibnamefont {Danailov}}, \bibinfo {author}
			{\bibfnamefont {Philipp~V.}\ \bibnamefont {Demekhin}}, \bibinfo {author}
			{\bibfnamefont {Michele~Di}\ \bibnamefont {Fraia}}, \bibinfo {author}
			{\bibfnamefont {Per}\ \bibnamefont {Eng-Johnsson}}, \bibinfo {author}
			{\bibfnamefont {Raimund}\ \bibnamefont {Feifel}}, \bibinfo {author}
			{\bibfnamefont {Guillaume}\ \bibnamefont {Gallician}}, \bibinfo {author}
			{\bibfnamefont {Luca}\ \bibnamefont {Giannessi}}, \bibinfo {author}
			{\bibfnamefont {Mathieu}\ \bibnamefont {Gisselbrecht}}, \bibinfo {author}
			{\bibfnamefont {Michele}\ \bibnamefont {Manfredda}}, \bibinfo {author}
			{\bibfnamefont {Michael}\ \bibnamefont {Meyer}}, \bibinfo {author}
			{\bibfnamefont {Catalin}\ \bibnamefont {Miron}}, \bibinfo {author}
			{\bibfnamefont {Jasper}\ \bibnamefont {Peschel}}, \bibinfo {author}
			{\bibfnamefont {Oksana}\ \bibnamefont {Plekan}}, \bibinfo {author}
			{\bibfnamefont {Kevin~C.}\ \bibnamefont {Prince}}, \bibinfo {author}
			{\bibfnamefont {Richard~J.}\ \bibnamefont {Squibb}}, \bibinfo {author}
			{\bibfnamefont {Marco}\ \bibnamefont {Zangrando}}, \bibinfo {author}
			{\bibfnamefont {Felipe}\ \bibnamefont {Zapata}}, \bibinfo {author}
			{\bibfnamefont {Shiyang}\ \bibnamefont {Zhong}}, \ and\ \bibinfo {author}
			{\bibfnamefont {Jan~Marcus}\ \bibnamefont {Dahlström}},\ }\bibfield  {title}
		{\enquote {\bibinfo {title} {Generation of entanglement using a
					short-wavelength seeded free-electron laser},}\ }\href {\doibase
			10.1126/sciadv.ado0668} {\bibfield  {journal} {\bibinfo  {journal} {Science
					Advances}\ }\textbf {\bibinfo {volume} {10}},\ \bibinfo {pages} {eado0668}
			(\bibinfo {year} {2024})}\BibitemShut {NoStop}%
		\bibitem [{\citenamefont {Umarova}\ \emph {et~al.}(2024)\citenamefont
			{Umarova}, \citenamefont {Umarov}, \citenamefont {T{\'o}th},\ and\
			\citenamefont {Csehi}}]{Umarova2024}%
		\BibitemOpen
		\bibfield  {author} {\bibinfo {author} {\bibfnamefont {Dilfuza}\ \bibnamefont
				{Umarova}}, \bibinfo {author} {\bibfnamefont {Otabek}\ \bibnamefont
				{Umarov}}, \bibinfo {author} {\bibfnamefont {Attila}\ \bibnamefont
				{T{\'o}th}}, \ and\ \bibinfo {author} {\bibfnamefont {Andr{\'a}s}\
				\bibnamefont {Csehi}},\ }\bibfield  {title} {\enquote {\bibinfo {title}
				{Spectral evidence of vibronic rabi oscillations in the resonance-enhanced
					photodissociation of mgh+},}\ }\href {\doibase 10.1103/PhysRevA.110.033112}
		{\bibfield  {journal} {\bibinfo  {journal} {Physical Review A}\ }\textbf
			{\bibinfo {volume} {110}},\ \bibinfo {pages} {033112} (\bibinfo {year}
			{2024})}\BibitemShut {NoStop}%
		\bibitem [{\citenamefont {de~las Heras}\ \emph {et~al.}(2025)\citenamefont
			{de~las Heras}, \citenamefont {Hern{\'a}ndez-Garc{\'i}a}, \citenamefont
			{Serrano}, \citenamefont {Prodanov}, \citenamefont {Popmintchev},
			\citenamefont {Popmintchev},\ and\ \citenamefont {Plaja}}]{deLasHeras2025}%
		\BibitemOpen
		\bibfield  {author} {\bibinfo {author} {\bibfnamefont {Alba}\ \bibnamefont
				{de~las Heras}}, \bibinfo {author} {\bibfnamefont {Carlos}\ \bibnamefont
				{Hern{\'a}ndez-Garc{\'i}a}}, \bibinfo {author} {\bibfnamefont {Javier}\
				\bibnamefont {Serrano}}, \bibinfo {author} {\bibfnamefont {Aleksandar}\
				\bibnamefont {Prodanov}}, \bibinfo {author} {\bibfnamefont {Dimitar}\
				\bibnamefont {Popmintchev}}, \bibinfo {author} {\bibfnamefont {Tenio}\
				\bibnamefont {Popmintchev}}, \ and\ \bibinfo {author} {\bibfnamefont {Luis}\
				\bibnamefont {Plaja}},\ }\bibfield  {title} {\enquote {\bibinfo {title}
				{Attosecond rabi oscillations in high harmonic generation resonantly driven
					by extreme ultraviolet laser fields},}\ }\href {\doibase
			10.1103/PhysRevResearch.7.023268} {\bibfield  {journal} {\bibinfo  {journal}
				{Physical Review Research}\ }\textbf {\bibinfo {volume} {7}},\ \bibinfo
			{pages} {023268} (\bibinfo {year} {2025})}\BibitemShut {NoStop}%
		\bibitem [{\citenamefont {Stammer}\ \emph {et~al.}(2023)\citenamefont
			{Stammer}, \citenamefont {Rivera-Dean}, \citenamefont {Maxwell},
			\citenamefont {Lamprou}, \citenamefont {Ord{\'o}{\~n}ez}, \citenamefont
			{Ciappina}, \citenamefont {Tzallas},\ and\ \citenamefont
			{Lewenstein}}]{Stammer2023QEDIntenseLM}%
		\BibitemOpen
		\bibfield  {author} {\bibinfo {author} {\bibfnamefont {Philipp}\ \bibnamefont
				{Stammer}}, \bibinfo {author} {\bibfnamefont {Javier}\ \bibnamefont
				{Rivera-Dean}}, \bibinfo {author} {\bibfnamefont {Andrew}\ \bibnamefont
				{Maxwell}}, \bibinfo {author} {\bibfnamefont {Theocharis}\ \bibnamefont
				{Lamprou}}, \bibinfo {author} {\bibfnamefont {Andr{\'e}s}\ \bibnamefont
				{Ord{\'o}{\~n}ez}}, \bibinfo {author} {\bibfnamefont {Marcelo~F}\
				\bibnamefont {Ciappina}}, \bibinfo {author} {\bibfnamefont {Paraskevas}\
				\bibnamefont {Tzallas}}, \ and\ \bibinfo {author} {\bibfnamefont {Maciej}\
				\bibnamefont {Lewenstein}},\ }\bibfield  {title} {\enquote {\bibinfo {title}
				{Quantum electrodynamics of intense laser-matter interactions: A tool for
					quantum state engineering},}\ }\href {\doibase 10.1103/PRXQuantum.4.010201}
		{\bibfield  {journal} {\bibinfo  {journal} {PRX Quantum}\ }\textbf {\bibinfo
				{volume} {4}},\ \bibinfo {pages} {010201} (\bibinfo {year}
			{2023})}\BibitemShut {NoStop}%
		\bibitem [{\citenamefont {de-la Peña}\ \emph {et~al.}(2025)\citenamefont
			{de-la Peña}, \citenamefont {Neufeld}, \citenamefont {Even~Tzur},
			\citenamefont {Cohen}, \citenamefont {Appel},\ and\ \citenamefont
			{Rubio}}]{delaPena2025QEDHHG}%
		\BibitemOpen
		\bibfield  {author} {\bibinfo {author} {\bibfnamefont {Sebastián}\
				\bibnamefont {de-la Peña}}, \bibinfo {author} {\bibfnamefont {Ofer}\
				\bibnamefont {Neufeld}}, \bibinfo {author} {\bibfnamefont {Matan}\
				\bibnamefont {Even~Tzur}}, \bibinfo {author} {\bibfnamefont {Oren}\
				\bibnamefont {Cohen}}, \bibinfo {author} {\bibfnamefont {Heiko}\ \bibnamefont
				{Appel}}, \ and\ \bibinfo {author} {\bibfnamefont {Angel}\ \bibnamefont
				{Rubio}},\ }\bibfield  {title} {\enquote {\bibinfo {title} {Quantum
					electrodynamics in high-harmonic generation: Multitrajectory ehrenfest and
					exact quantum analysis},}\ }\href {\doibase 10.1021/acs.jctc.4c01206}
		{\bibfield  {journal} {\bibinfo  {journal} {Journal of Chemical Theory and
					Computation}\ }\textbf {\bibinfo {volume} {21}},\ \bibinfo {pages} {283--290}
			(\bibinfo {year} {2025})}\BibitemShut {NoStop}%
		\bibitem [{SM()}]{SM}%
		\BibitemOpen
		\href@noop {} {\ }\bibinfo {note} {See Supplemental Material for details of
			the QED framework, the extended JC model, and the analytical derivations of
			the ATI-order-dependent AT splitting and yield scaling during Rabi
			oscillations.}\BibitemShut {Stop}%
		\bibitem [{\citenamefont {Manceau}\ \emph {et~al.}(2019)\citenamefont
			{Manceau}, \citenamefont {Spasibko}, \citenamefont {Leuchs}, \citenamefont
			{Filip},\ and\ \citenamefont {Chekhova}}]{Manceau2019IndefiniteMean}%
		\BibitemOpen
		\bibfield  {author} {\bibinfo {author} {\bibfnamefont {Mathieu}\ \bibnamefont
				{Manceau}}, \bibinfo {author} {\bibfnamefont {Kirill~Yu}\ \bibnamefont
				{Spasibko}}, \bibinfo {author} {\bibfnamefont {Gerd}\ \bibnamefont {Leuchs}},
			\bibinfo {author} {\bibfnamefont {Radim}\ \bibnamefont {Filip}}, \ and\
			\bibinfo {author} {\bibfnamefont {Maria~V}\ \bibnamefont {Chekhova}},\
		}\bibfield  {title} {\enquote {\bibinfo {title} {Indefinite-mean pareto
					photon distribution from amplified quantum noise},}\ }\href {\doibase
			10.1103/PhysRevLett.123.123606} {\bibfield  {journal} {\bibinfo  {journal}
				{Physical Review Letters}\ }\textbf {\bibinfo {volume} {123}},\ \bibinfo
			{pages} {123606} (\bibinfo {year} {2019})}\BibitemShut {NoStop}%
		\bibitem [{\citenamefont {Scully}\ and\ \citenamefont
			{Zubairy}(1997)}]{ScullyZubairy1997}%
		\BibitemOpen
		\bibfield  {author} {\bibinfo {author} {\bibfnamefont {Marlan~O.}\
				\bibnamefont {Scully}}\ and\ \bibinfo {author} {\bibfnamefont {M.~Suhail}\
				\bibnamefont {Zubairy}},\ }\href {\doibase 10.1017/CBO9780511813993} {\emph
			{\bibinfo {title} {Quantum Optics}}}\ (\bibinfo  {publisher} {Cambridge
			University Press},\ \bibinfo {address} {Cambridge},\ \bibinfo {year}
		{1997})\BibitemShut {NoStop}%
	\end{thebibliography}
\end{document}